\documentclass[useAMS,usenatbib]{mnras}

\usepackage{graphicx}	
\usepackage{subcaption}
\usepackage{floatrow}
\usepackage{amsmath}	
\usepackage{amssymb}	
\usepackage{multicol}        
\usepackage{bm}		

\newcommand{\dif}{\mathrm{d}}
\newcommand{\yr}{\,\mathrm{yr}}
\newcommand{\Myr}{\,\mathrm{Myr}}

\newcommand{\AU}{\,\mathrm{au}}

\newcommand{\RE}{R_{\oplus}}
\newcommand{\MSol}{M_{\odot}}
\newcommand{\RSol}{R_{\odot}}

\newcommand{\phihat}{\hat{\boldsymbol{\phi}}}
\newcommand{\uhat}{\hat{\mathbf{u}}}

\newcommand{\ehat}{\hat{\mathbf{e}}}
\newcommand{\qhat}{\hat{\mathbf{q}}}
\newcommand{\nhat}{\hat{\mathbf{n}}}
\newcommand{\lhat}{\hat{\boldsymbol{\ell}}}

\title[Planetesimals around polluted WDs]{High-eccentricity migration of planetesimals around polluted white dwarfs}

\author[O'Connor \& Lai]
{Christopher E. O'Connor\thanks{E-mail: coconnor@astro.cornell.edu} and Dong Lai \\
Department of Astronomy and Cornell Center for Astrophysics and Planetary Science, Cornell University, Ithaca, NY 14853, U.S.A.}

\begin{document}

\date{Accepted 2020 August 24. Received 2020 August 6; in original form 2020 May 11.}

\pagerange{\pageref{firstpage}--\pageref{lastpage}} \pubyear{2020}

\maketitle

\label{firstpage}

\begin{abstract}
    Several white dwarfs with atmospheric metal pollution have been found to host small planetary bodies (planetesimals) orbiting near the tidal disruption radius. We study the physical properties and dynamical origin of these bodies under the hypothesis that they underwent high-eccentricity migration from initial distances of several astronomical units. We examine two plausible mechanisms for orbital migration and circularization: tidal friction and ram-pressure drag in a compact disc. For each mechanism, we derive general analytic expressions for the evolution of the orbit that can be rescaled for various situations. We identify the physical parameters that determine whether a planetesimal's orbit can circularize within the appropriate time-scale and constrain these parameters based on the properties of the observed systems. For tidal migration to work, an internal viscosity similar to that of molten rock is required, and this may be naturally produced by tidal heating. For disc migration to operate, a minimal column density of the disc is implied; the inferred total disc mass is consistent with estimates of the total mass of metals accreted by polluted WDs.
\end{abstract}


\section{Introduction}

White dwarfs (WDs) frequently display the signatures of remnant planetary systems in the forms of atmospheric metal pollution and circumstellar debris \citep{Farihi2016}. The pollution rate among solitary WDs is between 25 and 50 per cent \citep{Zuckerman+2003,Zuckerman+2010,KGF2014}. A small fraction (less than $5$ per cent) also exhibits infrared excess, indicative of circumstellar dust or debris discs in close proximity to the WD \citep{FJZ2009,Barber+2012}. These phenomena are thought to originate from the tidal disruption and subsequent accretion of asteroids or minor planets that were excited onto highly eccentric orbits by companions to the WD \citep{Jura2003}, such as surviving planets \citep[e.g.,][]{DS2002,FH2014,Mustill+2018} or stellar binary partners \citep*[e.g.,][]{BV2015,PM2017,SNZ2017}. Observations of WD pollution thus allow insight into the structure and chemical composition of extrasolar rocky planets and small bodies \citep{Zuckerman+2007,Klein+2011,Xu+2019b,Bonsor+2020}.

The study of polluted WDs and their putative planetary systems has been enriched in recent years by the discovery of candidate planetary bodies in a few systems, all of which are apparently in states of ongoing disruption and accretion \citep{Vanderburg+2015,Manser+2019,Vanderbosch+2019,Gansicke+2019}. \citet{Vanderburg+2015} reported that the \textit{K2} light curve of WD~1145+017 (hereafter WD1145) contains several deep, asymmetric, transit-like signals with periods of $4.5$ to $5 \, {\rm h}$; they attributed these to at least one planetesimal ejecting small, rapidly disintegrating fragments (see also \citealt{Gansicke+2016}; \citealt{Rappaport+2016,Rappaport+2018}; \citealt{Xu+2019a}). In the case of SDSS~J1228+1040 (hereafter J1228), the presence of a planetesimal was inferred from the variability of Ca\,{\sc ii} emission from a known circumstellar gas disc on time-scales ranging from hours to decades: \citet{Manser+2019} identified the signal with a 2-h period with the object's Keplerian orbit. They further suggested that, if the orbit is moderately eccentric, relativistic apsidal precession could account for the observed variability of the source over decades \citep{Manser+2016}.

If these observations have been interpreted correctly, then there arise several questions about the nature and origin of these planetesimals. Their close proximity to the host WDs renders them vulnerable both to tidal disruption and to sublimation, yet in at least two cases the planetesimals have survived long enough to be observed. To some extent, this constrains the properties of the objects themselves: \citet{Manser+2019} argue that the object around J1228 must have at least the bulk density and rigidity of iron in order to endure the star's tidal gravity and must be at least several km in diameter in order to withstand UV radiation from the WD over several years.

The dynamical history of these objects is likewise puzzling, especially as it relates to the tidal-disruption theory of WD pollution. The prevailing view is that rocky bodies that enter the WD's tidal radius are completely disrupted, forming debris streams that circularize near the tidal radius on relatively short time-scales \citep[e.g.][]{Jura2003,Veras+2014,Veras+2015}. A possible observation of an extended debris stream transiting a polluted WD helps to corroborate this idea \citep{Vanderbosch+2019}. However, the candidate objects around WD1145 and J1228 apparently do not conform to this simplistic narrative. They are intact (or partly so), which suggests that they are monolithic bodies with internal cohesion rather than self-gravitating rubble piles. They also have quasi-circular orbits, which suggests that they have undergone some form of high-eccentricity migration from their original orbits beyond $1 \AU$. We will show that the conditions under which this migration can reproduce the orbits of these planetesimals place meaningful constraints on their properties and those of the WD's accretion system.

In this study, we examine two mechanisms that could facilitate high-eccentricity migration of small bodies into close orbits around WDs. In Section \ref{s:tides}, we consider orbital evolution due to static tides raised on a rocky planetesimal by the WD's gravity. In Section \ref{s:drag}, we explore evolution under drag forces in cases where the WD possesses an accretion disc. In Section \ref{s:discuss}, we discuss the implications of each scenario regarding the properties and origin of the candidate planetesimals. We review our main results in Section \ref{s:summary}.

\section{Tidal Migration} \label{s:tides}

Studies of high-eccentricity migration in planetary systems frequently invoke tidal dissipation as a means to shrink and circularize planetary orbits. Previous works have focussed on giant planets and mostly adopted the theory of weak tidal friction. In this section, we study the migration of small bodies into tight orbits under internal tidal dissipation (hereafter ``tidal migration''). We first present a general formalism to compare different models of dissipation; we then apply that formalism to two dissipation models and extract constraints on the parameters of these models from the properties of the candidate planetesimals around WDs.

\subsection{Formalism} \label{s:tides:formalism}

Consider a small body in orbit around a star of mass $M_{*}$. Let the orbital semi-major axis be $a$ and eccentricity $e$. We characterize the body by a linear size $s \ll a$ and bulk density $\rho$, from which we estimate its mass $M_{\rm b} = \rho s^{3} \ll M_{*}$ and moment of inertia $I_{\rm b} = \rho s^{5}$.

The response of an extended body to tidal forcing can be expressed in general as an infinite series of Fourier-like oscillatory components. The resulting expressions for the energy transfer rate and tidal torque are \citep[e.g.,][]{SL2014}
\begin{subequations} \label{eq:tides_Edot_Ldot}
\begin{align}
    \dot{E} &= \frac{5}{4 \pi} n E_{0} \sum_{m,N} m [W_{2m} F_{mN}(e)]^{2} \Im(\tilde{k}_{2}^{mN}), \\
    \dot{L} &= \frac{5}{4 \pi} E_{0} \sum_{m,N} N [W_{2m} F_{mN}(e)]^{2} \Im(\tilde{k}_{2}^{mN}),
\end{align}
\end{subequations}
where the series include terms with $m \in \{ 0, \pm 2 \}$ and $N$ any integer. Note $W_{2,0} = -(\pi / 5)^{1/2}$ and $W_{2,\pm 2} = (3 \pi / 10)^{1/2}$. The quantity $\tilde{k}_{2}^{mN}$ is the complex-valued Love number, to be discussed later in this section; $\Im(\tilde{k}_{2}^{mN})$ is its imaginary part. The orbital frequency (or mean motion) is $n = (G M_{*} / a^{3})^{1/2}$ and the characteristic tidal energy is given by
\begin{equation}
    E_{0} = \frac{G M_{*}^{2} s^{5}}{a^{6}}.
\end{equation}
In equations (\ref{eq:tides_Edot_Ldot}), the Hansen coefficient is given by
\begin{equation}
    F_{mN}(e) = \frac{1}{\pi} \int_{0}^{\pi} \frac{\cos\left[  m \psi(\xi) - N \phi(\xi) \right]}{(1 - e \cos\xi)^{2}} \, \dif\xi,
\end{equation}
where the eccentric anomaly $\xi$ is related to the mean and true anomalies $\phi$ and $\psi$ according to
\begin{subequations}
\begin{align}
    \phi &= \xi - e \sin\xi, \\
    \cos\psi &= \frac{\cos\xi - e}{1 - e \cos\xi}.
\end{align}
\end{subequations}

The evolution of the orbital elements $a$ and $e$ is given by
\begin{subequations}
\begin{align}
    \frac{1}{a} \frac{\dif a}{\dif t} &= \frac{\dot{E}}{|E|}, \\
    \frac{1}{e} \frac{\dif e}{\dif t} &= \frac{1 - e^{2}}{2 e^{2}} \left[ \frac{\dot{E}}{|E|} - \frac{2 \dot{L}}{L} \right],
\end{align}
\end{subequations}
where the energy and angular momentum of the orbit (assuming $M_{\rm b} \ll M_{*}$) are
\begin{subequations}
\begin{align}
    E &= - \frac{G M_{*} M_{\rm b}}{2 a}, \\
    L &= M_{\rm b} \left[ G M_{*} a \left( 1 - e^{2} \right) \right]^{1/2}.
\end{align}
\end{subequations}
By the conservation of total angular momentum, the body's rotation rate $\Omega_{\rm s}$ evolves according to
\begin{equation}
    \frac{1}{\Omega_{\rm s}} \frac{\dif \Omega_{\rm s}}{\dif t} = - \frac{\dot{L}}{I_{\rm b} \Omega_{\rm s}},
\end{equation}
assuming that the body has zero obliquity.

In this formalism, the physical processes that give rise to tidal dissipation are encoded by the complex Love number $\tilde{k}_{2}$, more specifically its imaginary part $\Im(\tilde{k}_{2})$. In general, $\tilde{k}_{2}$ is a function of the forcing frequency $\omega$; because each component $(m,N)$ of the tidal response has a different forcing frequency
\begin{equation}
    \omega_{mN} \equiv Nn - m \Omega_{\rm s},
\end{equation}
each has a distinct Love number $\tilde{k}_{2}^{mN} \equiv \tilde{k}_{2}(\omega_{mN})$. These quantities constitute the principal source of uncertainty in studying the tidal evolution of astrophysical systems, as they depend both on the internal structure of the dissipating body and the specific microphysical process responsible for dissipation. In the remainder of this section, we employ two tidal models to illustrate the tidal evolution of a rocky planetesimal around a WD.

To compare the overall efficiency of tidal dissipation between different models, we define functions $Z_{1}$ and $Z_{2}$ and rewrite equations (\ref{eq:tides_Edot_Ldot}) as:
\begin{subequations} \label{eq:tides_Z1Z2_def}
\begin{align}
    \dot{E} &= - \frac{n E_{0}}{(1-e)^{6}} Z_{1}(e, \Omega_{\rm s}/\Omega_{\rm p}), \label{eq:tides_Z1_def} \\
    \dot{L} &= - \frac{E_{0}}{(1-e)^{9/2}} Z_{2}(e, \Omega_{\rm s}/\Omega_{\rm p}),
\end{align}
\end{subequations}
where $\Omega_{\rm s}$ is the dissipating body's rotation rate and
\begin{equation}
    \Omega_{\rm p} = n \left[ \frac{1+e}{(1-e)^{3}} \right]^{1/2}
\end{equation}
is its orbital angular velocity at pericenter. These functions are defined in the same spirit as the functions $F_{E}$ and $F_{T}$ used by \citet{VL2019} but differ by factors related to the Love numbers and eccentricity. The characteristic time-scale of tidal migration is related to $Z_{1}$:
\begin{subequations} \label{eq:tides_ta}
\begin{align}
    t_{a} &\equiv \left| \frac{a}{\dot{a}} \right| = \frac{\rho r_{\rm p}^{6}}{2 s^{2}} \left( \frac{a}{G M_{*}^{3}} \right)^{1/2} | Z_{1}(e, \Omega_{\rm s}/\Omega_{\rm p}) |^{-1}, \\
    &\approx \frac{30 \Myr}{|Z_{1}|} \left( \frac{r_{\rm p}}{\RSol} \right)^{6} \left( \frac{s}{1 \, {\rm km}} \right)^{-2} \left( \frac{M_{*}}{\MSol} \right)^{-3/2} \nonumber \\
    & \hspace{1.5cm} \times \left( \frac{\rho}{1 \, {\rm g \, cm^{-3}}} \right) \left( \frac{a}{1 \AU} \right)^{1/2}, \label{eq:ta_numbers}
\end{align}
\end{subequations}
where $r_{\rm p} = a (1-e)$ is the separation between the body and the star at pericenter. The time-scale of spin--orbit synchronization is similarly related to $Z_{2}$.

The dissipation functions $Z_{1}$ and $Z_{2}$ are sensitive to several quantities. We have expressed them explicitly as functions of the eccentricity of a body's orbit and of its rotation rate normalized by a characteristic orbital frequency. Implicitly, but equally importantly, they depend also on several parameters that characterize the tidal dissipation mechanism under consideration, namely those that appear in the function $\tilde{k}_{2}(\omega)$.

\subsubsection{Weak Friction} \label{s:tides:formalism:weak}

Studies of tidal dissipation in planetary systems frequently adopt the theory of weak tidal friction \citep[e.g.][]{Alexander1973,Hut1981}, in which the tidal bulge raised on a body lags behind the orbit by a small, constant time $\tau$. The ansatz for the imaginary part of the complex Love number (for all $m$, $N$) is
\begin{equation}
    \Im(\tilde{k}_{2}) = k_{2} \omega \tau,
\end{equation}
where $k_{2}$ (the real-valued Love number) and $\tau$ are independent of $m$ and $N$ or the forcing frequency $\omega$. We relate the lag-time to the tidal quality factor
\begin{equation}
    Q \equiv \frac{1}{\Omega_{\rm p} \tau}.
\end{equation}
The summations over the forcing components in equations (\ref{eq:tides_Edot_Ldot}) yield closed-form expressions for $Z_{1}$ and $Z_{2}$:
\begin{subequations} \label{eq:Z1Z2_weak}
\begin{align}
    Z_{1}(e,\Omega_{\rm s}/\Omega_{\rm p}) &= \frac{ 3k_{2}/Q }{(1+e)^{8}} \left[ p_{1}(e) - (1+e)^{2} p_{2}(e) \frac{\Omega_{\rm s}}{\Omega_{\rm p}} \right], \label{eq:Z1_weak} \\
    Z_{2}(e,\Omega_{\rm s}/\Omega_{\rm p}) &= \frac{ 3k_{2}/Q }{(1+e)^{13/2}} \left[ p_{2}(e) - (1+e)^{2} p_{5}(e) \frac{\Omega_{\rm s}}{\Omega_{\rm p}} \right], \label{eq:Z2_weak}
\end{align}
\end{subequations}
where
\begin{subequations}
\begin{align}
    p_{1}(e) &= 1 + \frac{31}{2} e^{2} + \frac{255}{8} e^{4} + \frac{185}{16} e^{6} + \frac{25}{64} e^{8}, \\
    p_{2}(e) &= 1 + \frac{15}{2} e^{2} + \frac{45}{8} e^{4} + \frac{5}{16} e^{6}, \\
    p_{5}(e) &= 1 + 3 e^{2} + \frac{3}{8} e^{4}.
\end{align}
\end{subequations}
One can see from equation (\ref{eq:Z2_weak}) that weak tidal friction drives a body into stable, pseudosynchronous rotation at a rate given by
\begin{equation} \label{eq:spin_PS}
    \left( \frac{\Omega_{\rm s}}{\Omega_{\rm p}} \right)_{\rm ps} = \frac{1}{(1+e)^{2}} \frac{p_{2}(e)}{p_{5}(e)}.
\end{equation}
Since $L \gg I_{\rm b} \Omega_{\rm s}$, this synchronization occurs rapidly compared to the evolution of $a$ and $e$. Substitution of equation (\ref{eq:spin_PS}) into equation (\ref{eq:Z1_weak}) shows that the tidal evolution of a weakly dissipating body in pseudosynchronous rotation is entirely determined by $e$ and $k_{2}/Q$:
\begin{equation}
    Z_{1}(e) = \frac{ 3k_{2}/Q }{(1+e)^{8}} \left\{ p_{1}(e) - \frac{[p_{2}(e)]^{2}}{p_{5}(e)} \right\}.
\end{equation}
For $e \ll 1$, we find $Z_{1} \simeq (21/2) (k_{2}/Q) e^{2}$; whilst for $e \to 1$, $Z_{1} \simeq 0.15 (k_{2}/Q)$. Thus the tidal migration time-scale (equation \ref{eq:ta_numbers}) for highly eccentric initial orbits is:
\begin{align}
    t_{a} &\approx \frac{200 \Myr}{k_{2}/Q} \left( \frac{r_{\rm p}}{\RSol} \right)^{6} \left( \frac{s}{1 \, {\rm km}} \right)^{-2} \left( \frac{M_{*}}{\MSol} \right)^{-3/2} \nonumber \\
    & \hspace{1.5cm} \times \left( \frac{\rho}{1 \, {\rm g \, cm^{-3}}} \right) \left( \frac{a}{1 \AU} \right)^{1/2}. \label{eq:ta_numbers_weak}
\end{align}

\subsubsection{Viscoelastic Dissipation} \label{s:tides:formalism:visco}

Tidal friction in a realistic rocky or icy body may be described by viscoelastic theory, in which material can behave like both an elastic solid and a viscous fluid, depending on the nature of the applied strain \citep{TS2002}. There exist many rheological models for viscoelastic substances. Following \citet{SL2014}, we adopt the Maxwell model for our discussion.

A Maxwell material has two rheological parameters: the shear modulus (or rigidity) $\mu$ and the viscosity $\eta$. The transition between elastic and viscous behaviours is characterized by the Maxwell frequency 
\begin{equation}
    \omega_{\rm M} \equiv \mu/\eta.
\end{equation}
When forced at frequencies $\omega \gg \omega_{\rm M}$, the material behaves like an elastic solid; at low frequencies $\omega \ll \omega_{\rm M}$, it behaves like a viscous fluid.

\citet{SL2014} calculated the complex Love number as a function of forcing frequency $\omega$ for a homogeneous, spherical body composed of viscoelastic rock. We adapt their result to the problem at hand simply by replacing the spherical radius with our characteristic size parameter $s$ and by estimating the object's surface gravity as $g \sim G M_{\rm b} / s^{2} = G \rho s$. For a Fourier component with frequency $\omega$, the imaginary part of the Love number is
\begin{equation} \label{eq:k2Q_VE}
    \Im(\tilde{k}_{2}) = \frac{57 \omega \eta}{4 \beta} \left[ 1 + \left( \frac{\omega \eta}{\mu} \right)^{2} \left( 1 + \frac{19 \mu}{2 \beta} \right)^{2} \right]^{-1},
\end{equation}
where $\beta \equiv G (\rho s)^{2}$ is a measure of the object's self-gravity with dimensions of stress or pressure. It is convenient to define dimensionless variables
\begin{equation}
    \bar{\omega} \equiv \frac{\omega}{\Omega_{\rm p}}, \hspace{0.25cm} \bar{\mu} \equiv \frac{19 \mu}{2 \beta}, \hspace{0.25cm} \bar{\eta} \equiv \frac{19 \Omega_{\rm p} \eta}{2 \beta}
\end{equation}
such that
\begin{equation} \label{eq:k2Q_VE_dmls}
    \Im(\tilde{k}_{2}) = \frac{3}{2} \bar{\omega} \bar{\eta} \left[ 1 + \left( \frac{\bar{\omega} \bar{\eta}}{\bar{\mu}} \right)^{2} (1 + \bar{\mu})^{2} \right]^{-1}.
\end{equation}
Using this model, we can calculate the tidal energy transfer rate and torque via equations (\ref{eq:tides_Edot_Ldot}).

The material and structural properties of a small body are unknown. Because extrasolar small bodies have mostly rocky compositions \citep{Xu+2019b}, we have some recourse to geophysical constraints and laboratory measurements. If we adopt the appropriate rigidity for silicate rock or solid iron at low temperatures and pressures, the dimensionless rigidity is
\begin{equation}
    \bar{\mu} = 1.2 \times 10^{9} \left( \frac{\mu}{500 \, {\rm kbar}} \right) \left( \frac{\rho}{2.5 \, {\rm g \, cm^{-3}}} \right)^{-2} \left( \frac{s}{1 \, {\rm km}} \right)^{-2}.
\end{equation}
Rocky bodies with sizes $s \lesssim 1000 \, {\rm km}$ satisfy $\bar{\mu} \gg 1$. In that limit, equation (\ref{eq:k2Q_VE_dmls}) reduces to
\begin{equation} \label{eq:k2Q_mN_mubig}
    \Im(\tilde{k}_{2}) \simeq \frac{3}{2} \frac{\bar{\omega} \bar{\eta}}{1 + ( \bar{\omega} \bar{\eta} )^{2}}.
\end{equation}
Evidently, the tidal dissipation in a small body is determined primarily by the dimensionless viscosity $\bar{\eta}$ \citep[see also][]{Efroimsky2015}. A nominal value for this quantity is
\begin{align}
    \bar{\eta} &\approx 3.8 \times 10^{9} \left( \frac{\eta}{1 \, {\rm bar \yr}} \right) \left( \frac{\rho}{1 \, {\rm g \, cm^{-3}}} \right)^{-2} \left( \frac{s}{1 \, {\rm km}} \right)^{-2} \nonumber \\
    & \hspace{1.5cm} \times \left( \frac{1-e}{10^{-3}} \right)^{-3/2} \left( \frac{P}{1 \yr} \right)^{-1}, \label{eq:etabar_scaling}
\end{align}
where $P$ is the orbital period and where $\eta = 1 \, {\rm bar \, yr}$ is a characteristic viscosity for water ice. Geophysical viscosities vary over many orders of magnitude, depending on the composition and state of the material in question.

We note that the tidal theory for a Maxwell body reduces to the weak friction theory when the series in equations (\ref{eq:tides_Edot_Ldot}) comprise only terms where the combination $\bar{\omega} \bar{\eta} \ll 1$. In that case, we have $\Im(\tilde{k}_{2}) \propto \omega$ as before.

\subsection{Behavior of the Dissipation Functions} \label{s:tides:Zstuff}

\begin{figure}
    \centering
    \includegraphics[height=2.5in]{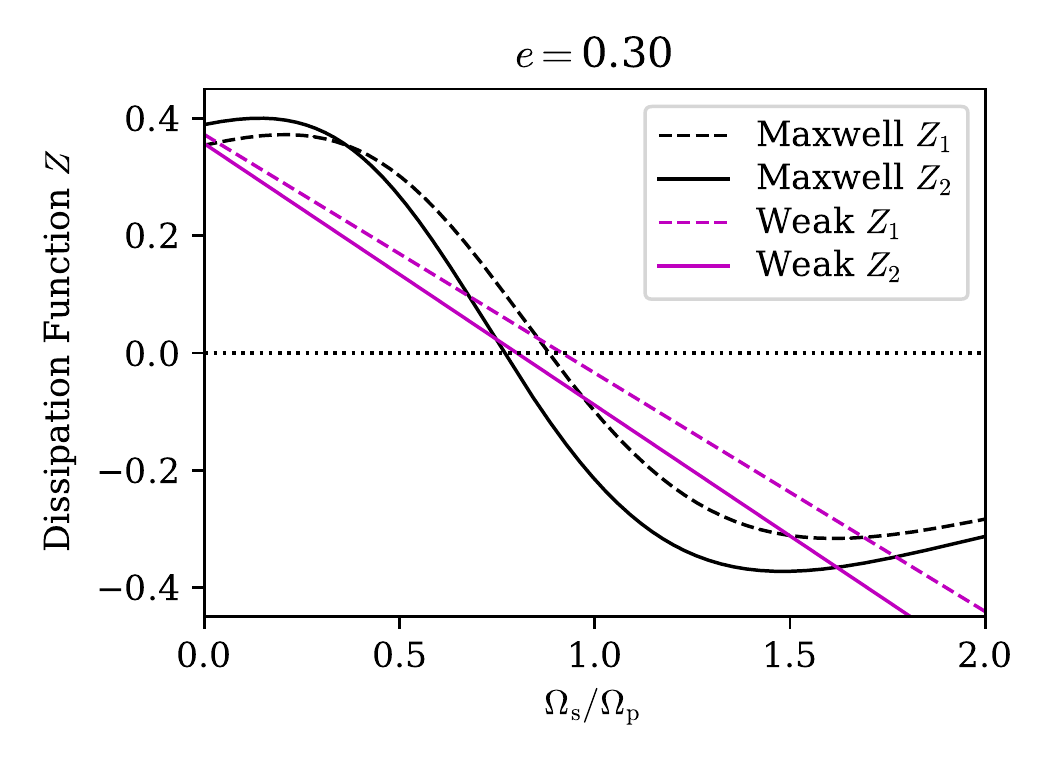}
    \includegraphics[height=2.5in]{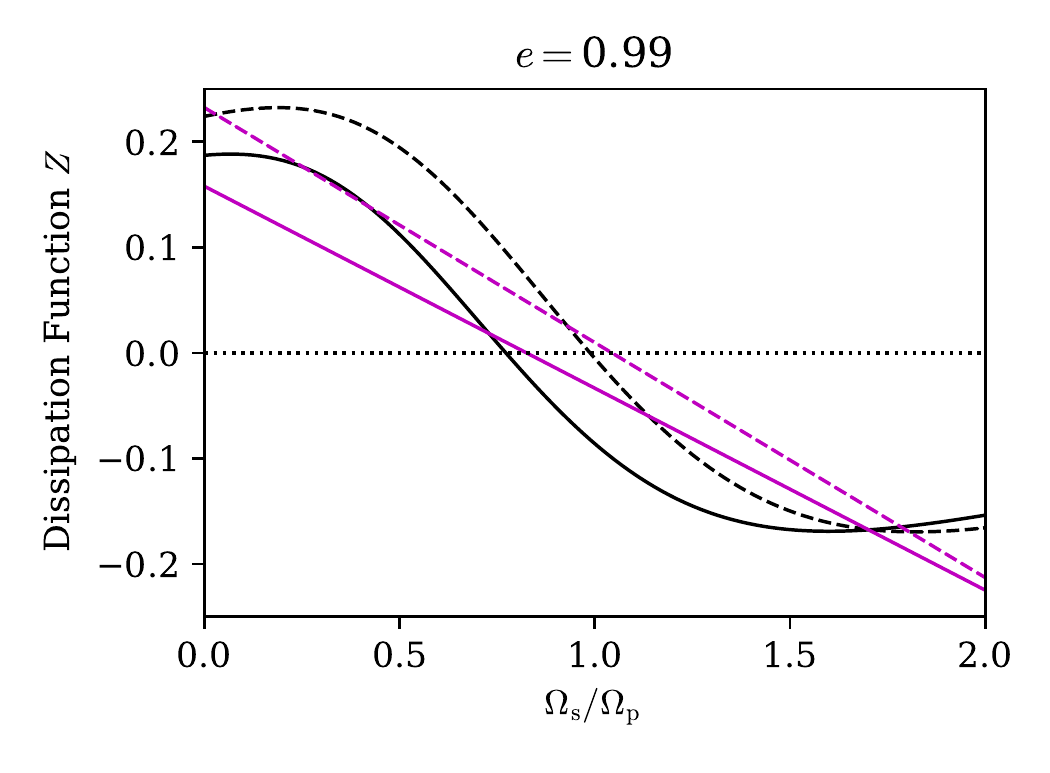}
    \caption{Upper panel: Tidal dissipation functions $Z_{1}$ (dashed curves) and $Z_{2}$ (solid), as defined in equations \ref{eq:tides_Edot_Ldot} and \ref{eq:tides_Z1Z2_def}, plotted with respect to the dimensionless rotation rate $\Omega_{\rm s}/\Omega_{\rm p}$ for an orbital eccentricity of $e = 0.3$. Black curves show the result of the Maxwell viscoelastic model with $\bar{\mu} \gg 1$ and $\bar{\eta} = 1$; magenta curves show the result of weak tidal friction with $k_{2}/Q = 1$. Lower panel: The same, but with an orbital eccentricity $e=0.99$.}
    \label{fig:Ztide_spin_lovisc}
\end{figure}

We now illustrate the properties of the dissipation functions $Z_{1}$ and $Z_{2}$ (defined through equations \ref{eq:tides_Edot_Ldot} and \ref{eq:tides_Z1Z2_def}) for both of our tidal models. In Figure \ref{fig:Ztide_spin_lovisc}, we show $Z_{1}$ and $Z_{2}$ with respect to $\Omega_{\rm s}/\Omega_{\rm p}$ for cases of moderate ($e=0.3$) and high ($0.99$) orbital eccentricities. For these examples, we have chosen $\bar{\eta} = 1$ and the value of $k_{2}/Q$ so that the functions have similar magnitude. We see that, in this instance, the Maxwell model and weak friction yield qualitatively similar results. The functions $Z_{2}$ both admit a stable, pseudosynchronous state for $\Omega_{\rm s}/\Omega_{\rm p} > 0$, albeit at slightly different rotation rates.

\begin{figure*}
    \centering
    \includegraphics[height=2.5in]{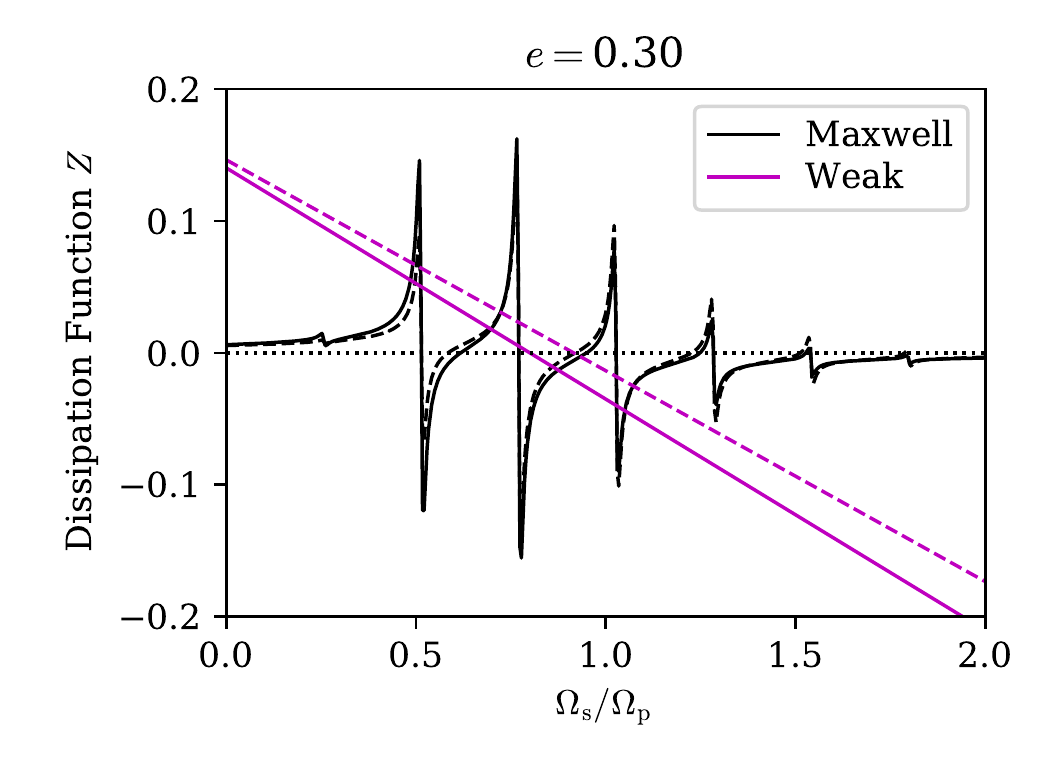}
    \includegraphics[height=2.5in]{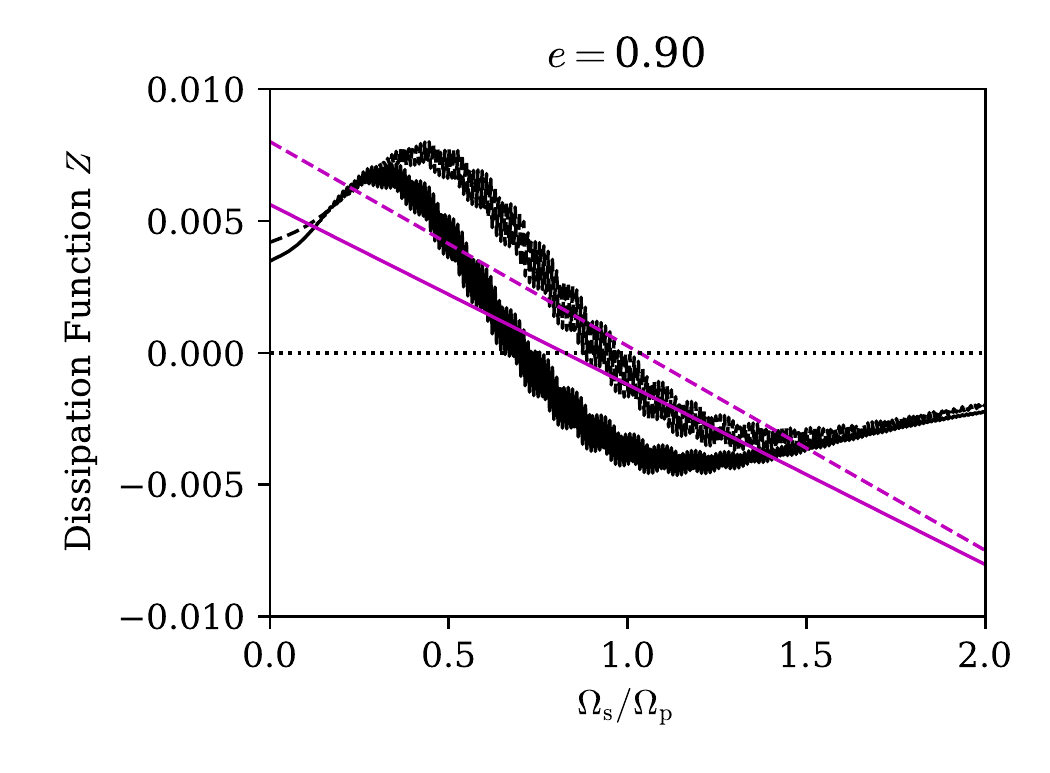}
    \includegraphics[height=2.5in]{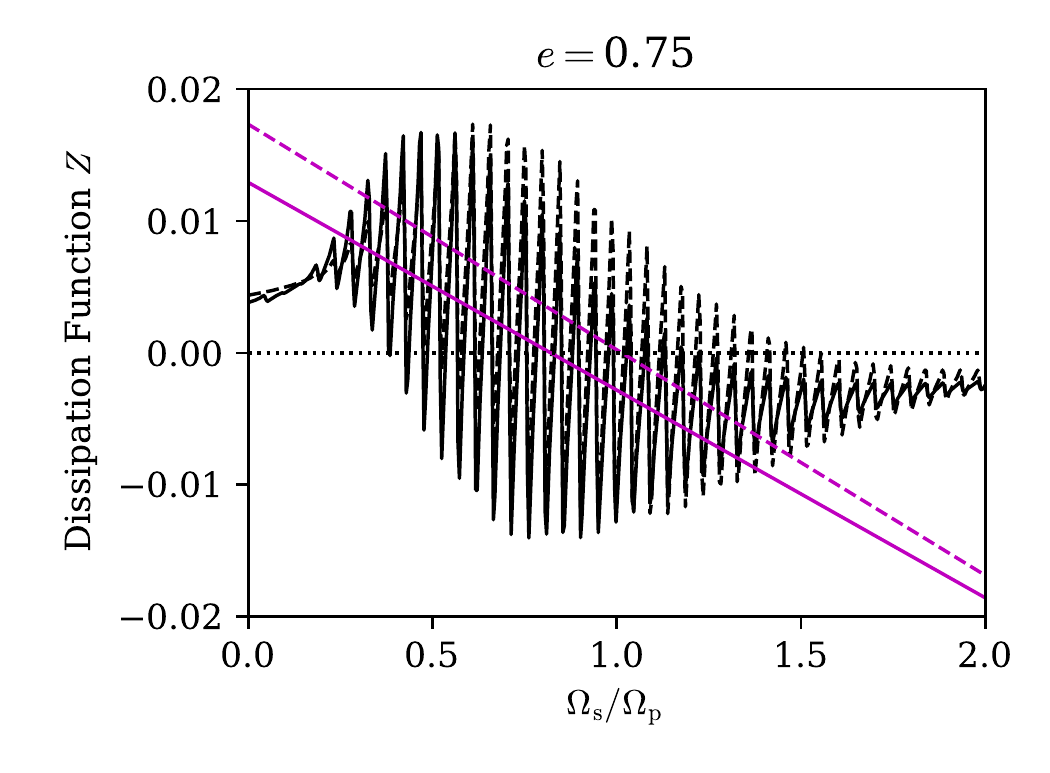}
    \includegraphics[height=2.5in]{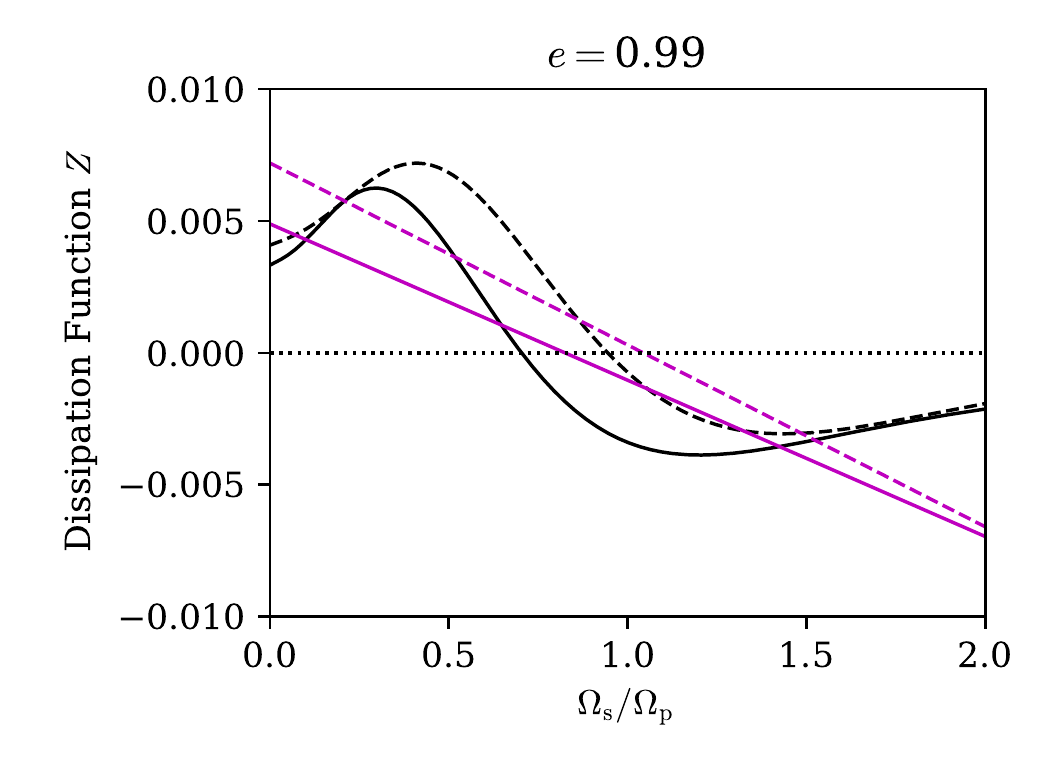}
    \caption{Each panel is similar to those of Fig.~\ref{fig:Ztide_spin_lovisc}, but with different choices of eccentricity and, for the Maxwell dissipation model, a higher dimensionless viscosity $\bar{\eta} = 10^{2}$. Note the transition from ``jagged'' behaviour at moderate eccentricities to ``smooth'' behaviour as $e \to 1$.}
    \label{fig:Ztide_spin_hivisc}
\end{figure*}

Figure \ref{fig:Ztide_spin_hivisc} shows $Z_{1}$ and $Z_{2}$ for a wide range of eccentricities and dissipation parameters. The qualitative behaviour of weak tidal friction is independent of $k_{2}/Q$ because the pseudosynchronous rotation rate is a function of eccentricity only and the pseudosynchronous state is always stable. This is not the case for the Maxwell model: in the limit where viscous stresses overwhelm a body's self-gravity ($\bar{\eta} \gg 1$, in our notation), the behaviour of these functions with respect to $\Omega_{\rm s}/\Omega_{\rm p}$ is quite rich. There are two main qualitative behaviours of $Z_{1}$ and $Z_{2}$ in this limit, to which we refer as ``jagged'' and ``smooth.'' Jagged behaviour is characterized by numerous zero-crossings, both stable and unstable; while smooth behaviour is characterized by a single, stable pseudosynchronous state. We explain the emergence of both behaviours in terms of the microphysics of dissipation in Appendix \ref{app:viscotides}. For the present discussion, it suffices to say that jagged behaviour occurs for dimensionless viscosity $\bar{\eta} \gtrsim \Omega_{\rm p}/n \sim (1-e)^{-3/2}$.

\begin{figure}[t]
    \centering
    \includegraphics[width=\columnwidth]{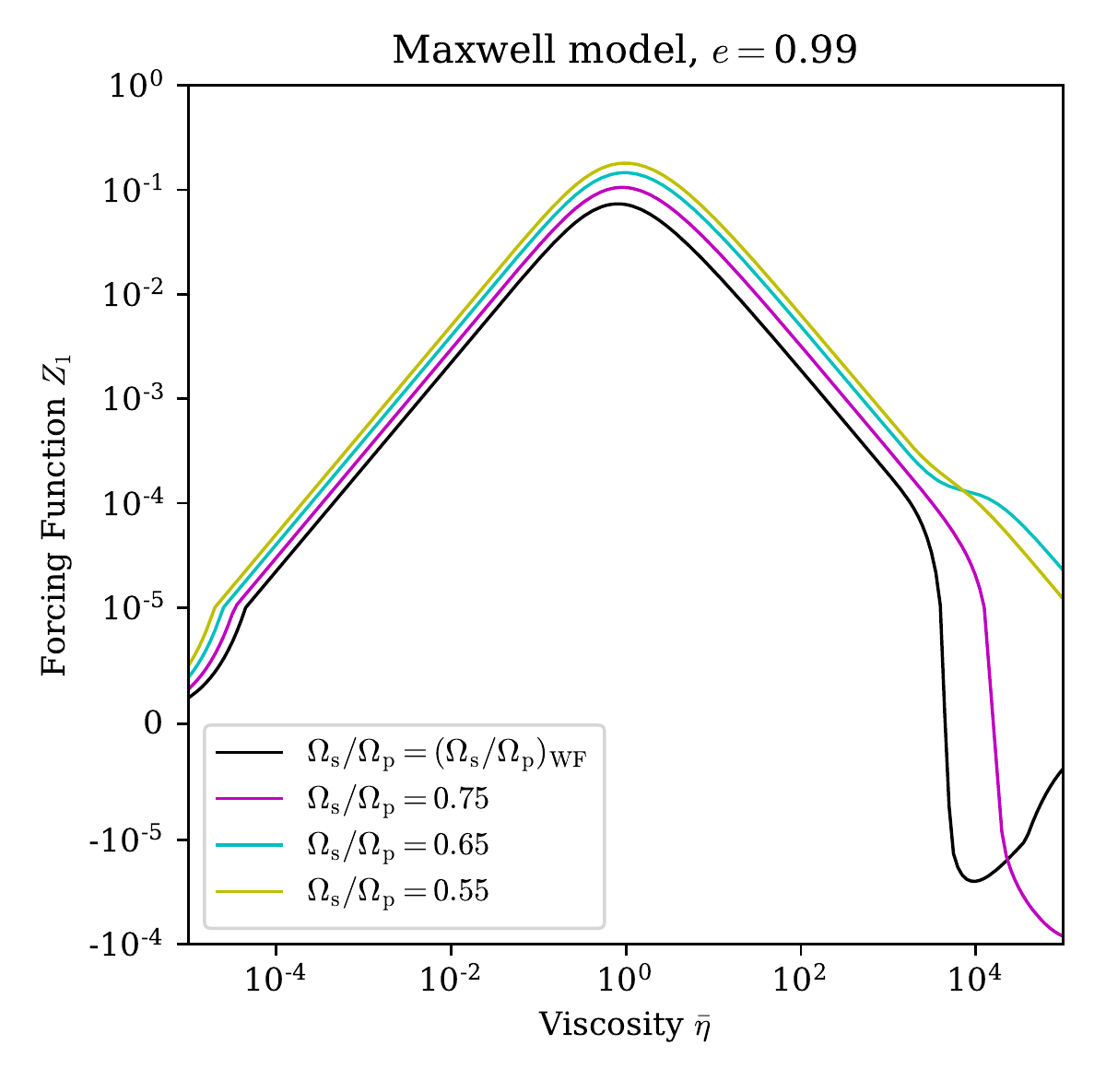}
    \caption{A bi-symmetric logarithmic plot of $Z_{1}$ versus $\bar{\eta}$ in the Maxwell model (with $\bar{\mu} \gg 1$) for $e = 0.99$ and various rotation rates.}
    \label{fig:Z1_visco}
\end{figure}

The complexity of the dissipation functions in the high-viscosity limit makes an analytic prescription for rotation elusive: the pseudosynchronous value of $\Omega_{\rm s} / \Omega_{\rm p}$ cannot be expressed in closed form as a function of $\bar{\eta}$ and $e$. Consequently, we must take a simpler tack to estimate the value of $Z_{1}$. In Figure \ref{fig:Z1_visco}, we display $Z_{1}$ as a function of $\bar{\eta}$ for a fixed eccentricity $e = 0.99$ and several rotation rates. For viscosity $\bar{\eta} \lesssim (1-e)^{-3/2}$, $Z_{1}$ is qualitatively insensitive to the assumed rotation rate. We identify three main regimes:
\begin{enumerate}
    \item[(i)] For $\bar{\eta} \ll 1$, we find $Z_{1} \propto \bar{\eta}$; this is consistent with convergence to weak tidal friction.
    \item[(ii)] A maximum $Z_{1} \sim 0.1$ occurs at $\bar{\eta} \sim 1$.
    \item[(iii)] For $1 \lesssim \bar{\eta} \lesssim (1-e)^{-3/2}$, we find $Z_{1} \propto \bar{\eta}^{-1}$. This corresponds to the quasi-elastic response of a Maxwell material.
\end{enumerate}
For extremely high viscosity $\bar{\eta} \gtrsim (1-e)^{-3/2}$, the behaviour of $Z_{1}$ is jagged and therefore highly spin-dependent. For some relatively rapid rotation rates, $Z_{1}$ abruptly shifts from positive to negative, meaning that energy is transferred from the spin to the orbit. For relatively slow rotation, $Z_{1}$ remains positive. Overall, the scaling $|Z_{1}| \propto \bar{\eta}^{-1}$ persists regardless of rotation. Thus, we conclude that tidal friction is most efficient when the dimensionless viscosity $\bar{\eta}$ is of the order of unity.

\begin{figure}[t]
    \centering
    \includegraphics[width=\columnwidth]{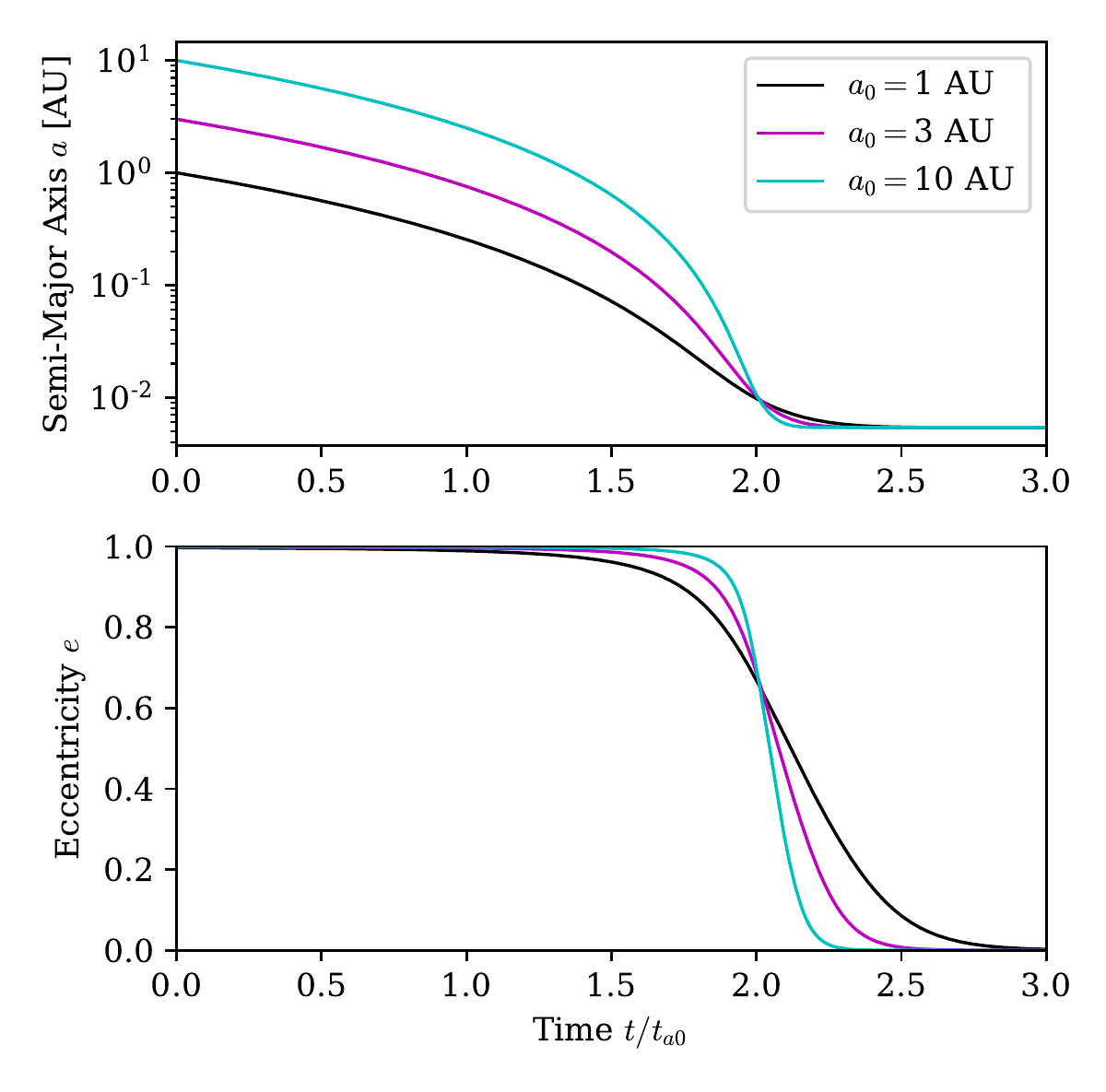}
    \caption{Three examples of the tidal evolution of planetesimals. The upper panel shows the evolution of the semi-major axis $a$ and the lower panel the eccentricity $e$. The time $t$ is normalized by $t_{a0}$, the timescale $t_{a}$ (see equation \ref{eq:tides_ta}) evaluated at the initial condition. All three examples start with the same pericentre distance $r_{\rm p} = 0.58 \RSol$, but different initial $a$ (as labelled). We use the weak-friction theory in these calculations, but similar results can be obtained using the Maxwell dissipation model. These examples show that $2.5 t_{a0}$ provides a good estimate for the tidal circularization time-scale $t_{\rm circ}$ (defined in the text). Note that  $t_{a0} \propto (a_{0})^{1/2}$; thus the true circularization time-scale is longer for $a_{0} = 10 \AU$ than for $a_{0} = 1 \AU$.}
    \label{fig:tides_calib}
\end{figure}

\subsection{Constraints on Tidal Migration} \label{s:tides:constraints}

Having established the properties and behaviour of the dissipation function $Z_{1}$, we are now able to evaluate the conditions under which tidal friction can mediate the high-eccentricity migration of planetesimals, such as those observed by \citet{Vanderburg+2015} and \citet{Manser+2019}. Given a tidal theory with specified values of the dissipation parameters (e.g., $k_{2}/Q$ for weak friction, $\bar{\mu}$ and $\bar{\eta}$ for viscoelastic friction), we assess the effectiveness of tidal friction by asking whether the WD's cooling age ($t_{\rm WD}$) is {\it longer} or {\it shorter} than the tidal circularization time-scale ($t_{\rm circ}$). We define $t_{\rm circ}$ as the time required for tidal evolution from an initial eccentricity $e \to 1$ to the value $e = 0.1$ (our results are insensitive to the exact choice of the lower eccentricity threshold). We have determined through numerical integration of the tidal evolution equations for a variety of initial conditions that the circularization time is typically between $2.2 t_{a0}$ and $2.6 t_{a0}$, where $t_{a0}$ is the value of equations (\ref{eq:tides_ta}) for the initial condition. Figure \ref{fig:tides_calib} depicts three such examples. Hereafter, we adopt the estimate $t_{\rm circ} \approx 2.5 t_{a0}$.

The time-scale $t_{a}$ depends strongly on the pericentre distance $r_{\rm p}$. To avoid tidal disruption, however, we require that $r_{\rm p}$ be greater than the critical radius for tidal disruption. For a body bound by self-gravity only, the tidal disruption (or Roche) radius is
\begin{equation}
    r_{\rm dis} = K R_{*} \left( \frac{\rho_{*}}{\rho} \right)^{1/3}.
\end{equation}
where $R_{*}$ and $\rho_{*}$ are the WD's radius and mean density. The numerical factor $K \sim 2$, depending on the body's shape, internal structure, and rotation. For a body with tensile strength $\gamma$ in addition to self-gravity, the generalized tidal disruption radius is
\begin{equation}
    r_{\rm dis} = K' R_{*} \left[ \frac{G \rho_{*} \rho s^{2}}{\gamma + \beta} \right]^{1/3},
\end{equation}
where again $\beta = G (\rho s)^{2}$ is a measure of self-gravity and where $K'$ is analogous to $K$. Self-gravity is the dominant binding force when $\bar{\gamma} \equiv \gamma/\beta \ll 1$. For the typical strength of terrestrial rock or iron,
\begin{equation}
    \bar{\gamma} \approx 1.5 \times 10^{6} \left( \frac{\gamma}{1 \, {\rm kbar}} \right) \left( \frac{\rho}{1 \, {\rm g \, cm^{-3}}} \right)^{-2} \left( \frac{s}{1 \, {\rm km}} \right)^{-2}.
\end{equation}
Thus, the disruption radius of a cohesive object can be much smaller than that of a rubble pile of the same size and bulk density. Since the Roche radius is typically $\sim \RSol$, km-sized objects with high tensile strengths may withstand disruption near the surface of a typical white dwarf ($\sim 10^{-2} \RSol$).

\subsubsection{WD~1145+017} \label{s:tides:constraints:WD1145}

WD1145 is transited by solid debris with orbital periods between $4.5$ and $4.9 \, {\rm h}$. The strongest transit signal occurs every $4.5 \, {\rm h}$ and exhibits highly variable depths \citep[e.g.,][]{Vanderburg+2015,Gansicke+2016,Rappaport+2016,Rappaport+2018,Croll+2017,Gary+2017}. This has been attributed to a disintegrating asteroid or planetesimal on a near-circular orbit with $a = 1.16 \RSol$ around the WD (given $M_{*} = 0.6 \MSol$). Following \citet{Rappaport+2016}, we take its mass to be $10^{23} \, {\rm g}$, roughly 10 per cent that of Ceres. We also assume a bulk density $\rho = 3 \, {\rm g \, cm^{-3}}$, which yields a size $s = 320 \, {\rm km}$.

Pseudosynchronous tidal migration conserves orbital angular momentum, meaning that the quantity $a (1-e^{2}) = r_{\rm p} (1+e)$ would have remained constant. This implies that the object's initial pericentre distance was half of its present semi-major axis, or about $0.58 \RSol$. To avoid disruption at this distance, the object's dimensionless tensile strength must be $\bar{\gamma} \gtrsim 7$; given the assumed size and density, its true tensile strength was greater than $\sim 4 \, {\rm kbar}$, which we note to be somewhat higher than the measured strengths of silicate meteorites \citep{Petrovic2001}. This suggests that the original object was a monolith rather than a rubble pile.

The cooling age of WD~1145+017 is some $200 \Myr$ \citep{Vanderburg+2015,Izquierdo+2018}. We take this as the upper limit of $t_{\rm circ}$ and therefore find $t_{a0} \lesssim 80 \Myr$. Assuming initial values $a = 5.0 \AU$ and $r_{\rm p} = 0.58 \RSol$, as well as the values of $\rho$, $s$, and $M_{*}$ mentioned above, we obtain a lower limit on the dissipation function $Z_{1}$ during migration, namely $Z_{1} \gtrsim 2.0 \times 10^{-6}$. For weak tidal friction, this corresponds to the quality factor $k_{2}/Q \gtrsim 1.3 \times 10^{-5}$ by the result $Z_{1} \simeq 0.15 (k_{2}/Q)$ for $e \to 1$. For Maxwell viscoelastic friction, we first note that the rigidity $\mu$ is presumably at least of the order of the minimal tensile strength required to avoid disruption, $\gamma \approx 4 \, {\rm kbar}$. This provides a lower bound of $\bar{\mu} \gtrsim 60$ on the dimensionless rigidity, which indicates that the approximation $\bar{\mu} \gg 1$ is reasonable. To constrain the viscosity, we use the asymptotic expressions we obtained for $\bar{\mu} \gg 1$ and $e \to 1$. In the low-viscosity limit, $Z_{1} \simeq 0.1 \bar{\eta}$ and we find $\bar{\eta} \gtrsim 2 \times 10^{-5}$. In the high-viscosity limit, $Z_{1} \simeq 0.1 / \bar{\eta}$ and thus $\bar{\eta} \lesssim 5 \times 10^{4}$.

The constraint on $Z_{1}$ is modified if we relax our assumed knowledge of one or more properties of the system. Because $r_{\rm p}$ is well constrained by conservation of angular momentum, $M_{*}$ is known from spectroscopic analysis of the WD, and $a$ has a relatively muted effect on $t_{a}$, the most important uncertainties are the values of $\rho$ and $s$. By treating these quantities as free parameters, we would find a less specific constraint
\begin{equation*}
    Z_{1} \geq 4.2 \times 10^{-5} \left( \frac{\rho}{3 \, {\rm g \, cm^{-3}}} \right) \left( \frac{s}{320 \, {\rm km}} \right)^{-2}.
\end{equation*}
For a larger or a less compact planetesimal, the corresponding constraint on $Z_{1}$ is weaker.

\subsubsection{SDSS~J1228+1040} \label{s:tides:constraints:J1228}

The candidate planetesimal around J1228 was discovered through spectroscopic rather than photometric observations. \citet{Manser+2019} used theoretical arguments to constrain the size of the hypothesized object within the range
\begin{equation*}
    4 \, {\rm km} \lesssim s \lesssim 600 \, {\rm km}.
\end{equation*}
In brief, the lower size limit follows from the requirement that radiation-driven sublimation of the body's surface is sufficient to provide the observed metal accretion rate. The upper limit follows from the requirement that the body withstand tidal disruption on its current orbit through internal strength alone. These calculations assumed the object's density to be $\rho \approx 8 \, {\rm g \, cm^{-3}}$, consistent with an iron-rich composition. However, the composition of the object is not constrained by existing observations, so as in Section \ref{s:tides:constraints:WD1145} we treat both the size and density as free parameters.

\citet{Manser+2019} calculated the planetesimal's orbital semi-major axis to be $a \approx 0.73 \RSol$. They also speculate that the orbit is moderately eccentric in order to explain the system's long-term variability relativistic precession; the required eccentricity for this is $e \approx 0.54$. We will consider the constraints on tidal migration that we obtain with and without the supposed eccentricity.

In the case without eccentricity, the reasoning used to constrain the value of $Z_{1}$ is the same as for WD1145. Using J1228's cooling age of $100 \Myr$ \citep{Gansicke+2006}, we find
\begin{equation*}
    Z_{1} \gtrsim 2 \times 10^{-4} \left( \frac{\rho}{8 \, {\rm g \, cm^{-3}}} \right) \left( \frac{s}{10 \, {\rm km}} \right)^{-2}
\end{equation*}
is required for tidal circularization. The corresponding constraints on the parameters of the tidal theory are
\begin{equation*}
    \frac{k_{2}}{Q} \gtrsim 1.4 \times 10^{-2} \left( \frac{\rho}{8 \, {\rm g \, cm^{-3}}} \right) \left( \frac{s}{10 \, {\rm km}} \right)^{-2}
\end{equation*}
for weak friction and
\begin{align*}
    \bar{\eta} &\gtrsim 2 \times 10^{-2} \left( \frac{\rho}{8 \, {\rm g \, cm^{-3}}} \right) \left( \frac{s}{10 \, {\rm km}} \right)^{-2}, \\
    \bar{\eta} &\lesssim 50 \left( \frac{\rho}{8 \, {\rm g \, cm^{-3}}} \right)^{-1} \left( \frac{s}{10 \, {\rm km}} \right)^{2}
\end{align*}
for viscoelastic friction. If the planetesimal's current orbit is eccentric, then the conserved quantity $a (1 - e^{2})$ would be smaller than a circular orbit would imply and so the initial pericentre distance would have been smaller as well. In effect, this loosens the constraint on the planetesimal's $Z_{1}$ by a factor of $(1-e^{2})^{6}$ because $t_{a} \propto r_{\rm p}^{6}$.

\section{Migration from Disc Drag} \label{s:drag}

The presence of accreting gaseous or dusty discs around some polluted WDs presents another possibility for high-eccentricity migration. A planetesimal would experience a drag force while passing through the circumstellar material, resulting in shrinkage and circularization of the orbit. Similar effects have been studied previously in different astrophysical contexts, such as active galactic nuclei \citep[e.g.,][]{Rauch1995,SK1999,ML2020} and protoplanetary discs \citep{Rein2012}. Indeed, \citet{GV2019} have discussed the possible relationship between compact accretion discs and close-in planetesimals around WDs. In this section, we present a simplified calculation of the disc migration scenario. Our formulation allows us to evaluate this migration mechanism efficiently for a variety of conditions and parameters.

Consider a small body moving with velocity $\mathbf{v}$ through a disc with local density $\rho_{\rm d}$. The disc itself rotates with bulk velocity $\mathbf{v}_{\rm d}$ about the central WD. It exerts a drag force on the body through ram pressure:
\begin{equation} \label{eq:Fdrag_tot}
    \mathbf{F} = - \frac{C}{2} \sigma \rho_{\rm d} | \mathbf{v} - \mathbf{v}_{\rm d} | ( \mathbf{v} - \mathbf{v}_{\rm d} ),
\end{equation}
where $\sigma$ is the body's cross-sectional area and $C$ is a dimensionless drag coefficient of order unity. As before, we define the body's size $s$ and bulk density $\rho$ so that its cross-section is $\sigma = s^{2}$ and its mass is $M_{\rm b} = \rho s^{3}$. Equation (\ref{eq:Fdrag_tot}) is valid for supersonic motion, when $\Delta v = |\mathbf{v} - \mathbf{v}_{\rm d}|$ is much larger than the local sound speed or the velocity dispersion of dust particles in the disc; or motion with high Reynolds number, ${\rm Re} = s \Delta v / \nu \gg 1$, where $\nu$ is the kinematic viscosity of the gas. These conditions are satisfied for thin discs (aspect ratio $h/r \ll 1$). We consider a Keplerian disc with constant aspect ratio $h/r \ll 1$ and a surface mass distribution $\Sigma_{\rm d}(r)$. We neglect any disc evolution due to accretion by the WD or perturbations by the encroaching body.

The body should experience a significant drag force only while it is within an altitude $z \sim h$ of the disc's midplane. For simplicity, we replace the local disc density $\rho_{\rm d}$ with its vertically averaged value $\Sigma_{\rm d} / 2 h$ and stipulate zero drag force for $|z| > h$. Thus the drag force per unit mass acting on the body while $|z| \leq h$ is
\begin{equation} \label{eq:fdrag_spec}
    \mathbf{f} = \frac{\mathbf{F}}{M_{\rm b}} = - \frac{C}{4 h} \frac{\Sigma_{\rm d}}{\rho s} | \Delta\mathbf{v} | \Delta\mathbf{v},
\end{equation}
where $\Delta\mathbf{v} = \mathbf{v} - \mathbf{v}_{\rm d}$.

The specific energy $\mathcal{E}$ and specific angular momentum $\ell$ of the small body are related to the orbital elements $a$ and $e$ by
\begin{subequations}
\begin{align}
    \mathcal{E} &= - \frac{1}{2} n^{2} a^{2}, \\
    \ell &= n a^{2} \left( 1 - e^{2} \right)^{1/2},
\end{align}
\end{subequations}
where $n = (G M_{*} / a^{3})^{1/2}$ is the mean motion. We denote the specific angular momentum vector $\boldsymbol{\ell} = \ell \lhat$ and the eccentricity vector $\mathbf{e} = e \ehat$. The relative orientation of the orbit and the disc is specified by the inclination angle $I$ and the apsidal angle $\varpi$, defined such that
\begin{equation} \label{eq:coordinate_vectors}
    \nhat = \lhat \cos I + \sin I \left( -\ehat \sin\varpi + \qhat \cos\varpi \right),
\end{equation}
where $\qhat = \lhat \times \ehat$ and $\nhat$ is the unit vector normal to the disc plane. We call $\varpi$ the ``apsidal'' angle in the sense that, when the body intersects the disc midplane such that $\mathbf{v} \cdot \nhat > 0$, it makes an angle $\varpi$ with respect to the reference direction $\ehat$.

If the orbit is inclined with respect to the disc midplane by an angle $I \gg h/r$, then the body experiences a nonzero drag force in short intervals near the line of nodes. In this case, the duration of disc-crossing is much shorter than the orbital period; thus, for orbits with $I \gg h/r$ we may use the impulse approximation to compute the orbital evolution. The aspect ratio of a WD's accretion disc is small, typically $\sim 10^{-3}$ for a gaseous disc and orders of magnitude smaller for dust grains \citep{Melis+2010}. Therefore, the impulse approximation is valid even for orbits with inclinations as small as $I \sim 0\fdg5$. We treat the orbital evolution for $I \lesssim h/r$, the ``coplanar'' limit, in Appendix \ref{app:coplanar}.

\subsection{Single Disc Passage} \label{s:drag:impulse}

Consider a single passage of the body through the disc. We approximate the its trajectory as a straight line traversed at constant velocity $\mathbf{v}$. We consider the region of the disc that the body crosses to be a homogeneous slab of vertical thickness $2 h$ and density $\Sigma_{\rm d} / 2 h$ that moves uniformly with the local Keplerian velocity $\mathbf{v}_{\rm d}$ directed in the plane normal to $\nhat$. The local properties of the disc and both velocity vectors are evaluated at the position $\mathbf{r}_{0}$, the point where the orbit intersects the disc midplane.

Under these assumptions, the duration of the particle's passage through the disc is
\begin{equation} \label{eq:Dt_drag}
    \Delta t = \frac{2 h}{| \mathbf{v} \cdot \nhat |}.
\end{equation}
The work done by the drag force on the orbit during a single passage is
\begin{equation} \label{eq:drag_dE}
    \delta\mathcal{E} = (\mathbf{f} \cdot \mathbf{v}) \Delta t = - \frac{C \Sigma_{\rm d}}{2 \rho s} \frac{|\Delta \mathbf{v}|}{|\mathbf{v} \cdot \nhat|} (\mathbf{v} \cdot \Delta\mathbf{v}).
\end{equation}
Note that $\delta\mathcal{E}$ does not depend explicitly on the disc's scale height $h$.

Similarly, the change of the body's angular momentum due to drag is
\begin{equation} \label{dlvec_drag}
    \delta\boldsymbol{\ell} = (\mathbf{r}_{0} \times \mathbf{f}) \Delta t = - \frac{C \Sigma_{\rm d}}{2 \rho s} \frac{ |\Delta\mathbf{v}|}{| \mathbf{v} \cdot \nhat |} (\mathbf{r}_{0} \times \Delta\mathbf{v}),
\end{equation}
which is also independent of $h$. The magnitude of the specific angular momentum vector $\boldsymbol{\ell}$ changes by
\begin{equation} \label{eq:drag_dlmag}
    \delta\ell = \delta\boldsymbol{\ell} \cdot \lhat
\end{equation}
and the unit vector $\lhat$ changes direction by
\begin{equation} \label{eq:drag_dlhat}
    \delta\lhat = \frac{\delta\boldsymbol{\ell}}{\ell} - \left( \frac{\delta\ell}{\ell} \right) \lhat = - \frac{C \Sigma_{\rm d}}{2 \rho s} \left( \frac{r_{0}}{\ell} \right)^{2} | \Delta\mathbf{v} | | \mathbf{v}_{\rm d} | (\uhat \times \lhat),
\end{equation}
where $\mathbf{r}_{0} = r_{0} \uhat$.

Finally, the change of the body's eccentricity vector can be calculated as
\begin{equation}
    \delta\mathbf{e} = \delta\left( \frac{\mathbf{v} \times \boldsymbol{\ell}}{G M_{*}} - \uhat \right) = \frac{(\delta\mathbf{v}) \times \boldsymbol{\ell} + \mathbf{v} \times (\delta\boldsymbol{\ell})}{G M_{*}},
\end{equation}
where we have neglected the change of $\uhat = \mathbf{r}_{0} / r_{0}$ according to the impulse approximation. As with the angular momentum, we can isolate the perturbation of the orientation of the unit vector $\ehat$ as
\begin{equation}
    \delta\ehat = \frac{\delta\mathbf{e}}{e} - \left( \frac{\delta e}{e} \right) \ehat
\end{equation}
where $\delta e \equiv \delta\mathbf{e} \cdot \ehat$.

\begin{figure}
    \centering
    \includegraphics[width=\columnwidth]{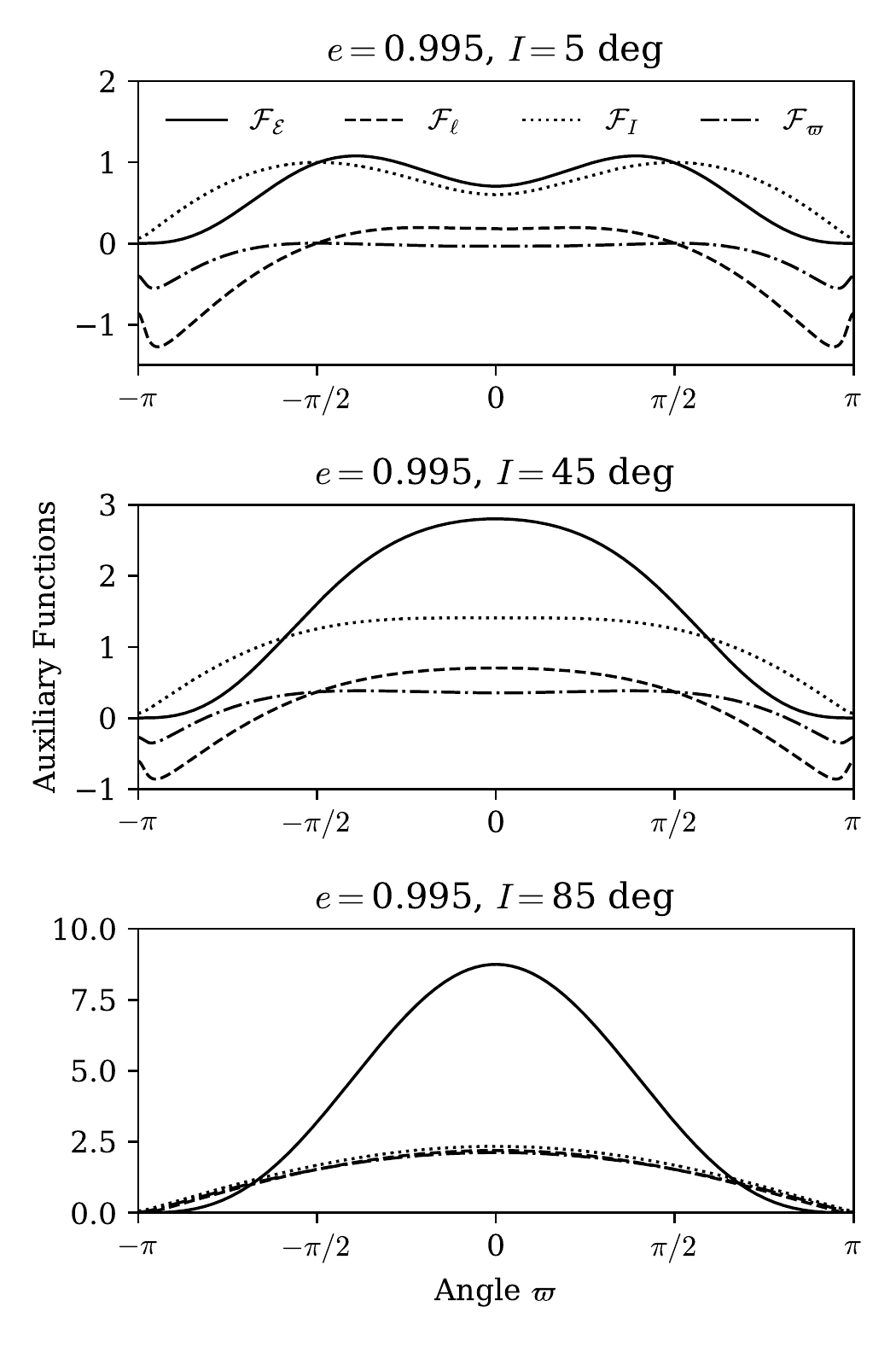}
    \caption{Auxiliary functions $\mathcal{F}_{\mathcal{E}}$,  $\mathcal{F}_{\ell}$, $\mathcal{F}_{I}$, and $\mathcal{F}_{\varpi}$ (see legend in top panel) from equations (\ref{eq:drag_dX/X}), plotted over the apsidal angle $\varpi$ for eccentricity $e = 0.995$ and inclinations of $I = 5 \, \deg$ (top panel), $45 \, \deg$ (middle), and $85 \, \deg$ (bottom).}
    \label{fig:auxF}
\end{figure}

The scalars $\delta\mathcal{E}$ and $\delta\ell$ and the vectors  $\delta\lhat$ and $\delta\ehat$ are sufficient to determine the evolution of the four orbital elements. As with tidal friction, the changes of $a$ and $e$ during a single disc passage are given by
\begin{subequations} \label{eq:drag_dOrb/Orb}
\begin{align}
    \frac{\delta a}{a} &= \frac{\delta \mathcal{E}}{|\mathcal{E}|}, \label{eq:drag_da/a} \\
    \frac{\delta e}{e} &= \frac{1 - e^{2}}{2 e^{2}} \left[ \frac{\delta\mathcal{E}}{\mathcal{|E|}} - 2 \left( \frac{\delta\ell}{\ell} \right) \right].
\end{align}
Using the identity $\cos I = \lhat \cdot \nhat$ and noting that the disc normal $\nhat$ is fixed (recall that we ignore the body's effect on the disc), we find the inclination angle changes by
\begin{equation}
    \delta I = - \frac{\delta\lhat \cdot \nhat}{\sin I}
\end{equation}
in a single passage through the disc. Indeed, equation (\ref{eq:drag_dlhat}) describes a small rotation of $\lhat$ by $\delta I$ about $\uhat$. Similarly, the change of the apsidal angle follows from the identity $\cos\varpi = \ehat \cdot \uhat$, which implies
\begin{equation} \label{eq:drag_dOm}
    \delta\varpi = - \frac{\delta\ehat \cdot \uhat}{\sin \varpi}
\end{equation}
\end{subequations}
For a Keplerian disc, we have
\begin{subequations} \label{eq:drag_dX/X}
\begin{align}
    \frac{\delta\mathcal{E}}{|\mathcal{E}|} &= - \frac{C \Sigma_{\rm d}}{\rho s} \frac{\mathcal{F}_{\mathcal{E}}(e,I,\varpi)}{(1-e^{2}) (1 + e \cos\varpi) \sin I}, \label{eq:drag_dE/E} \\
    \frac{\delta\ell}{\ell} &= - \frac{C \Sigma_{\rm d}}{2 \rho s} \frac{\mathcal{F}_{\ell}(e,I,\varpi)}{ (1 + e \cos\varpi) \sin I}, \label{eq:drag_dl/l} \\
    \delta I &= - \frac{C \Sigma_{\rm d}}{2 \rho s} \frac{\mathcal{F}_{I}(e,I,\varpi)}{(1 + e \cos\varpi)^{3/2}}, \label{eq:drag_dI} \\
    \delta\varpi &= - \frac{C \Sigma_{\rm d}}{\rho s} \frac{\mathcal{F}_{\varpi}(e,I,\varpi)}{e (1 + e \cos\varpi) \sin I} \label{eq:drag_dOm}
\end{align}
\end{subequations}
In equations (\ref{eq:drag_dX/X}), the functions $\mathcal{F}_{\mathcal{E}}$, $\mathcal{F}_{\ell}$, $\mathcal{F}_{I}$, and $\mathcal{F}_{\varpi}$ are dimensionless functions defined so as to be of order unity for all eccentricities and inclinations; explicit expressions for these functions can be obtained from equations (\ref{eq:drag_dE}) through (\ref{eq:drag_dOm}). In Figure \ref{fig:auxF}, we display these functions over the full range of $\varpi$ for various inclinations and a high eccentricity ($e = 0.995$).

Therefore, for a highly eccentric orbit (where $a/r_{0} \gg 1$), we find that $|\delta\mathcal{E} / \mathcal{E}| \gg |\delta\ell/\ell|$ and $|\delta I| \sim |\delta\ell/\ell|$. This implies that drag has a similar effect on the orbit to tidal friction in the high-eccentricity limit. The semi-major axis and eccentricity decrease with each orbit, the relative change of $e$ being slower than that of $a$ by a factor $\sim (1-e^{2})$. The inclination of the orbit decreases on a time-scale similar to that of eccentricity damping. The revolution of the line of apsides can be prograde or retrograde, depending on the value of $\mathcal{F}_{\varpi}$; however, we will show in Section \ref{s:drag:secular} that drag forces are not the dominant perturbation on $\varpi$ over many orbits and that the total precession rate is almost always negative.

\subsection{Secular Evolution} \label{s:drag:secular}

\begin{figure}
    \centering
    \includegraphics[width=\columnwidth]{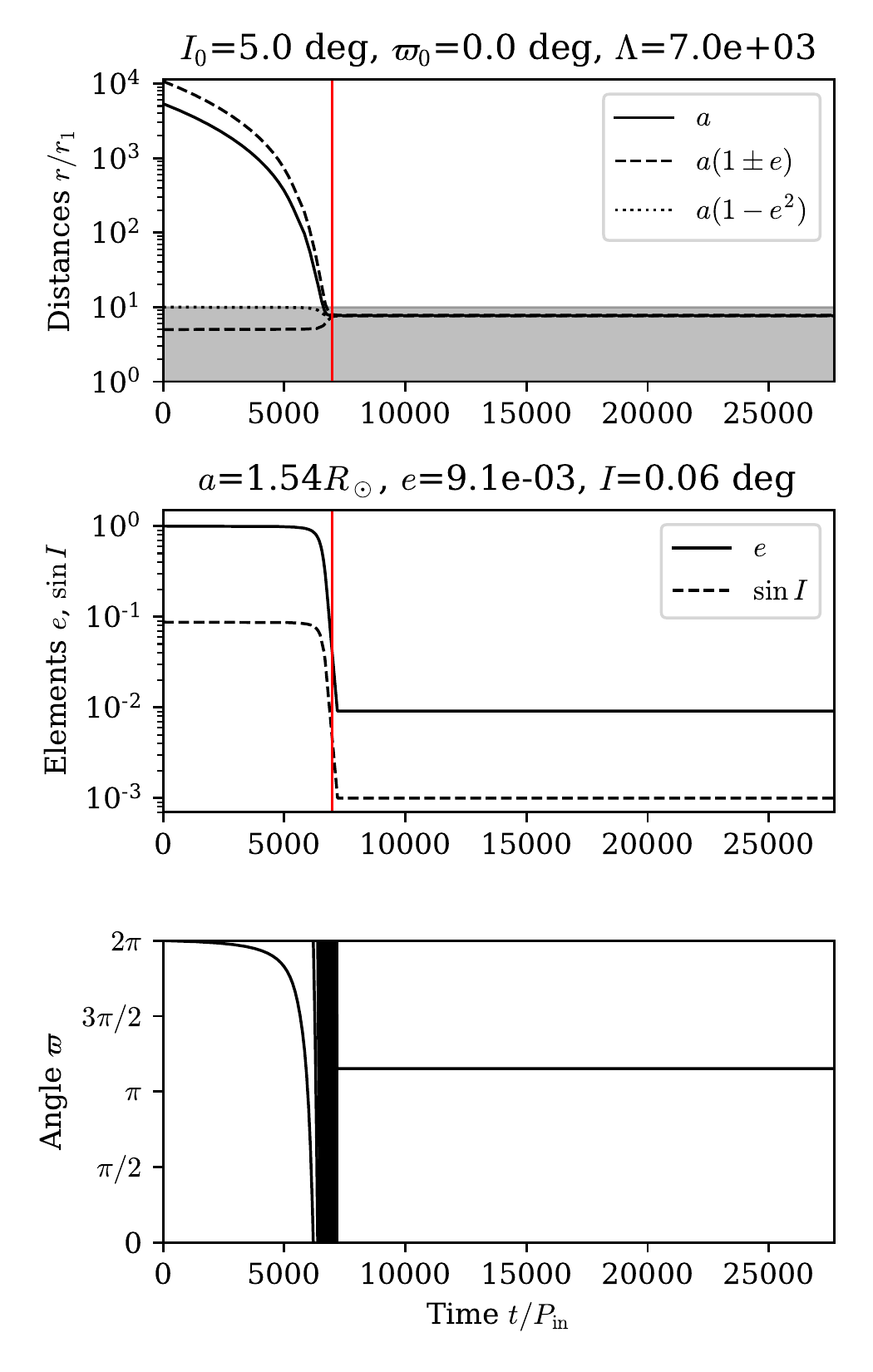}
    \caption{Secular evolution of a small body due to drag forces from repeated disc crossings. The migration parameter is $\Lambda \approx 7.0 \times 10^{3}$ and the initial orbital elements are $a = 5.0 \AU$, $r_{\rm p} = \RSol$, $I = 5^{\circ}$, and $\varpi = 0$. Time $t = \Lambda P_{\rm in}$ is marked with a vertical red line. Top panel: Characteristic orbital distances $a$ (solid black curve), $a (1 \pm e)$ (dashed), and $a (1-e^{2})$ (dotted) expressed in units of the disc's inner radius $r_{1}$. The shaded region indicates the radial extent of the circumstellar disc. Middle panel: Orbital elements $e$ (solid black) and $\sin I$ (dashed). The terminal values of $a$, $e$, and $I$ are listed above this panel. Bottom panel: Apsidal angle $\varpi$.}
    \label{fig:tilt_a5_I5_Om0_Lam7e3}
\end{figure}

\begin{figure}
    \centering
    \includegraphics[width=\columnwidth]{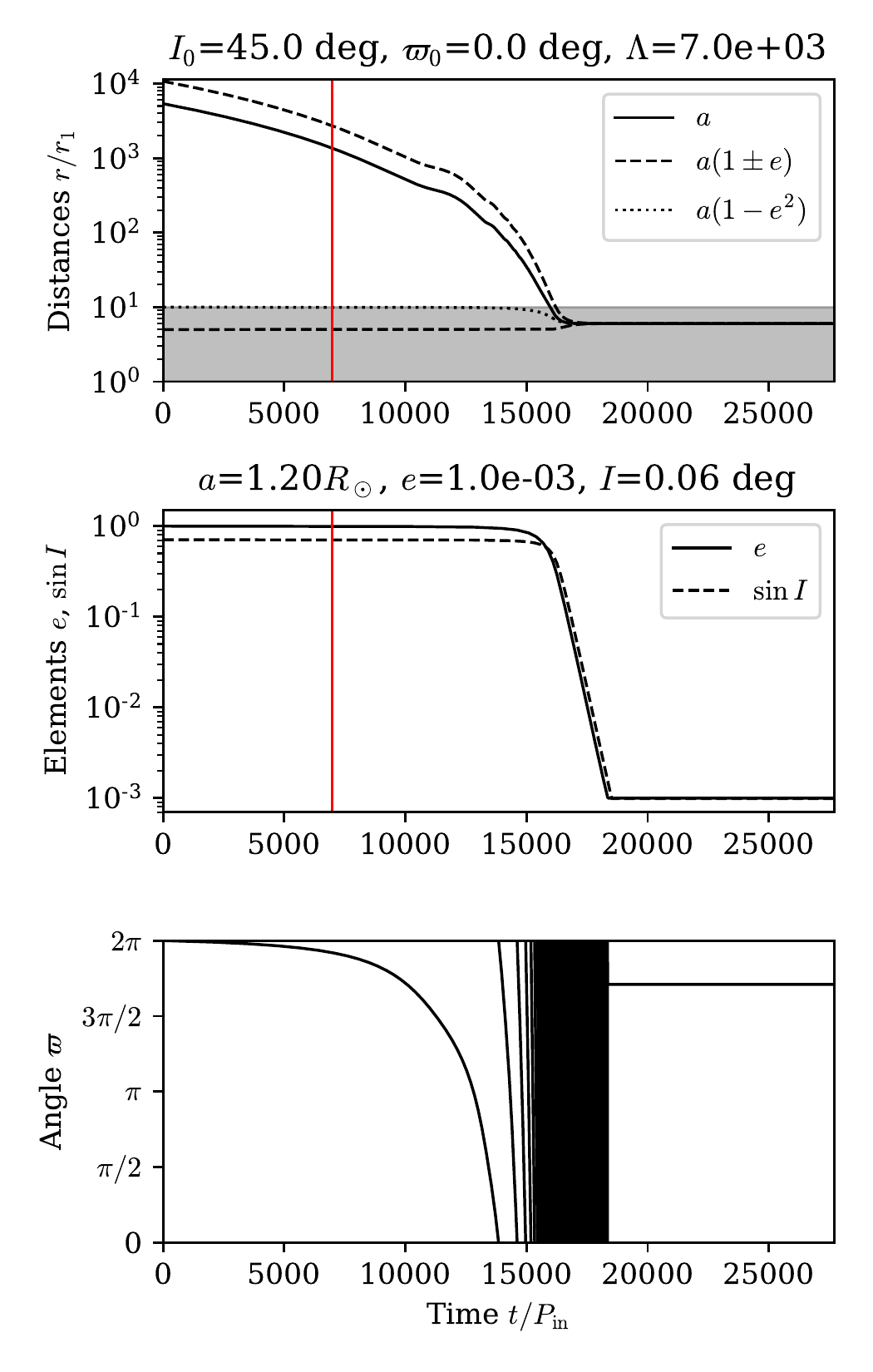}
    \caption{Same as Fig.~\ref{fig:tilt_a5_I5_Om0_Lam7e3} but with initial $I = 45^{\circ}$.}
    \label{fig:tilt_a5_I45_Om0_Lam7e3}
\end{figure}

\begin{figure}
    \centering
    \includegraphics[width=\columnwidth]{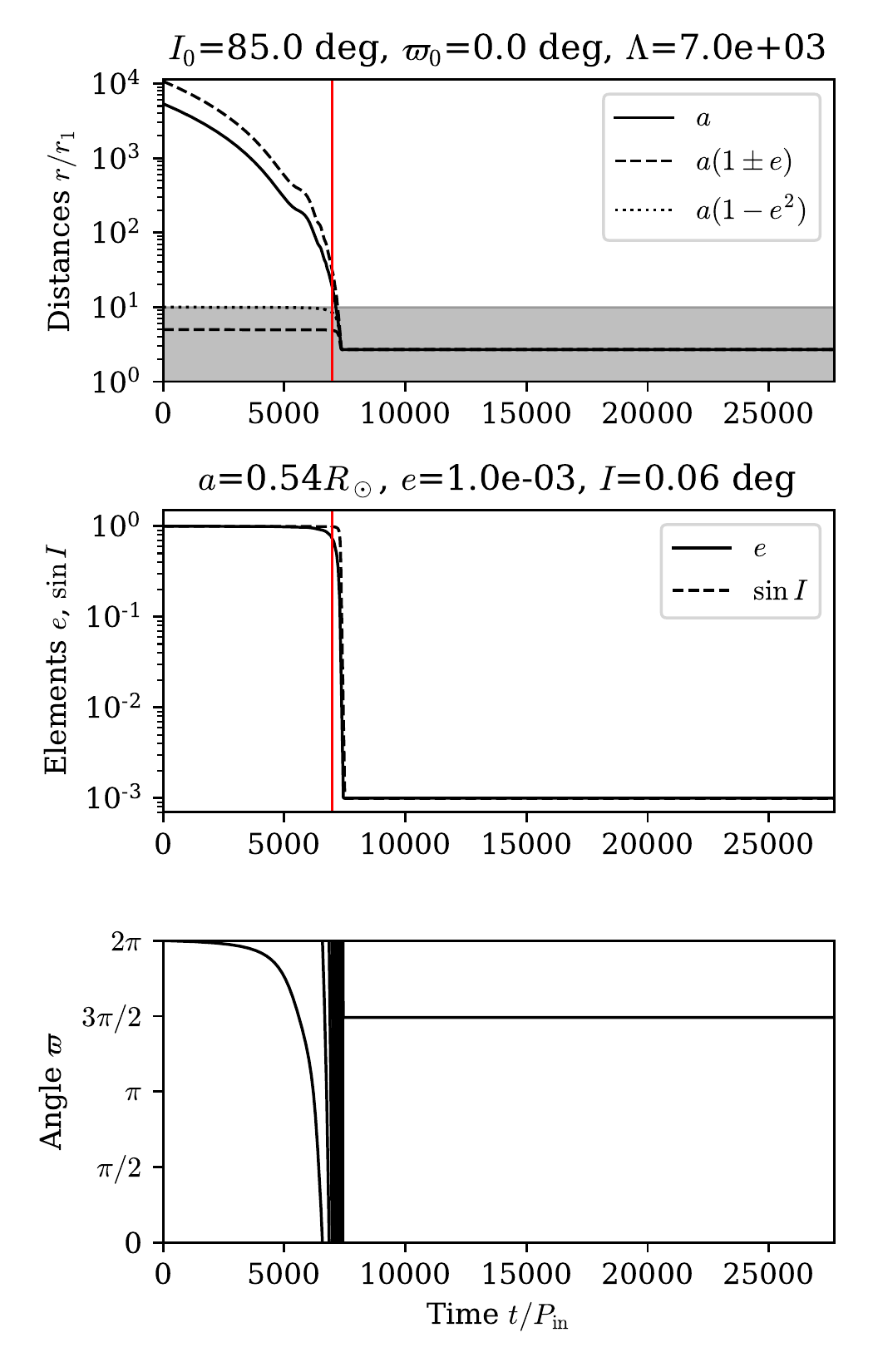}
    \caption{Same as Fig.~\ref{fig:tilt_a5_I5_Om0_Lam7e3} but with initial $I = 85^{\circ}$.}
    \label{fig:tilt_a5_I85_Om0_Lam7e3}
\end{figure}

\begin{figure}
    \centering
    \includegraphics[width=\columnwidth]{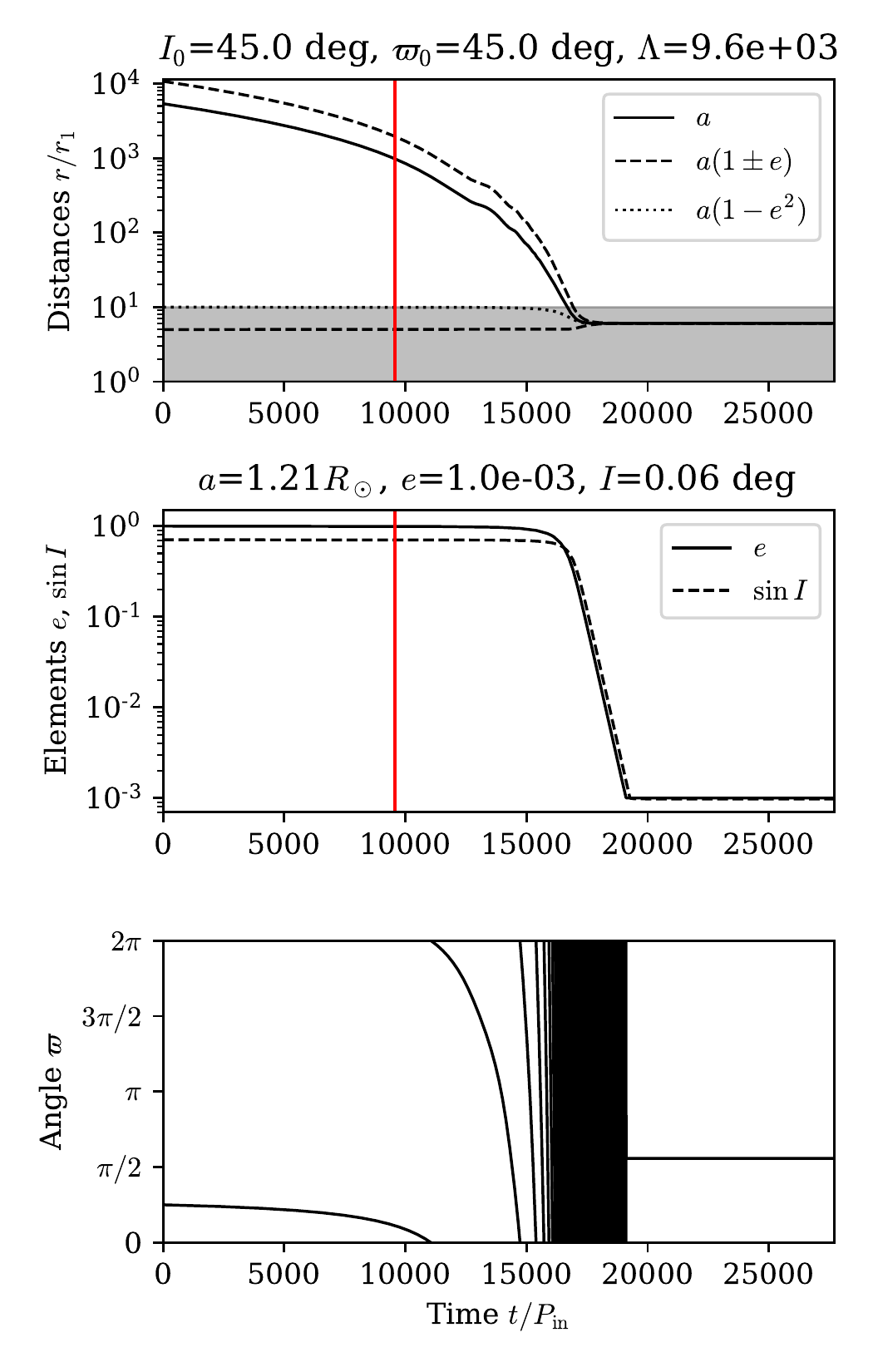}
    \caption{Same as Fig.~\ref{fig:tilt_a5_I45_Om0_Lam7e3} but with initial $\varpi = 45^{\circ}$ and $\Lambda \approx 1 \times 10^{4}$.}
    \label{fig:tilt_a5_I45_Om45_Lam1e4}
\end{figure}

\begin{figure}
    \centering
    \includegraphics[width=\columnwidth]{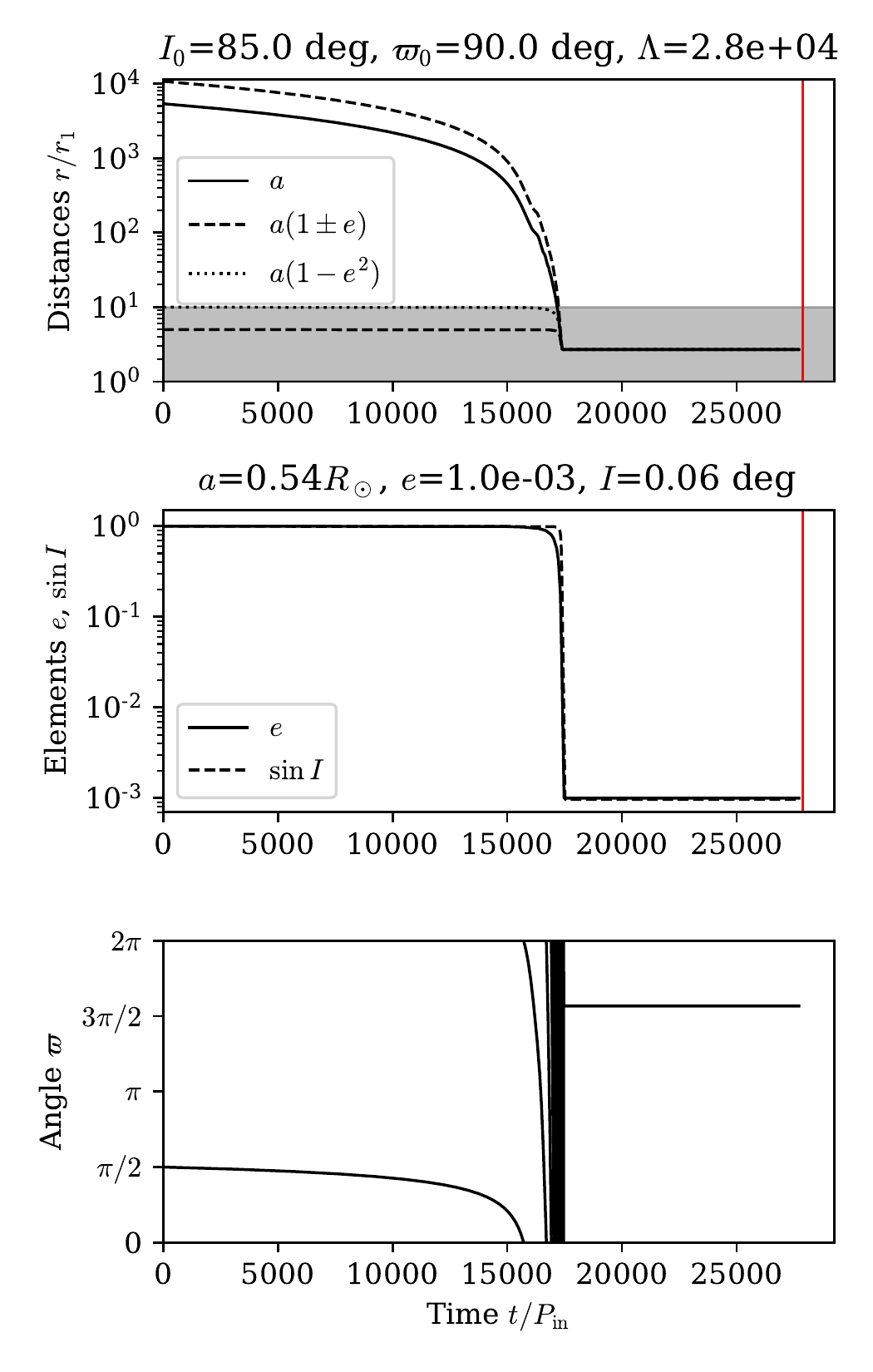}
    \caption{Same as Fig.~\ref{fig:tilt_a5_I85_Om0_Lam7e3} but with initial $\varpi = 90^{\circ}$ and $\Lambda \approx 3 \times 10^{4}$.}
    \label{fig:tilt_a5_I85_Om90_Lam3e4}
\end{figure}

It is simple to compute the secular evolution of the disc--orbit system under the impulse approximation. For a quasi-Keplerian orbit, we simply divide the total change accrued by the orbital elements over a possible two disc passages by the Keplerian period $P = 2 \pi / n$. Thus,
\begin{subequations}
\begin{align}
    \frac{\dif a}{\dif t} &= \frac{\delta a_{+} + \delta a_{-}}{P}, \label{eq:drag_dadt} \\
    \frac{\dif e}{\dif t} &= \frac{\delta e_{+} + \delta e_{-}}{P}, \\
    \frac{\dif I}{\dif t} &= \frac{\delta I_{+} + \delta I_{-}}{P}, \\
    \frac{\dif \varpi}{\dif t} &= \frac{\delta \varpi_{+} + \delta \varpi_{-}}{P}, \label{eq:dOmdt_drag}
\end{align}
\end{subequations}
where the subscripts $(+)$ and $(-)$ refer to the ascending and descending nodes, respectively.

From equations (\ref{eq:drag_da/a}), (\ref{eq:drag_dE/E}), and (\ref{eq:drag_dadt}), we find that the characteristic time-scale of migration due to disc drag is
\begin{equation} \label{eq:tdrag}
    t_{\rm drag} = \frac{\rho s}{C \Sigma_{0}} (1 - e_{\rm in}) P_{\rm in},
\end{equation}
where $P_{\rm in}$ and $e_{\rm in}$ are the initial period and eccentricity of the orbit and $\Sigma_{0} \equiv \Sigma_{\rm d}(r_{0})$. The characteristic number of orbits over which migration and circularization occur is
\begin{subequations} \label{eq:Lambda}
\begin{align}
    \Lambda &\equiv \frac{\rho s}{C \Sigma_{0}} (1 - e_{\rm in}), \\
    &= \frac{100}{C} \left( \frac{\rho}{1 \, {\rm g \, cm^{-3}}} \right) \left( \frac{s}{1 \, {\rm km}} \right) \left( \frac{\Sigma_{0}}{1 \, {\rm g \, cm^{-2}}} \right)^{-1} \left( \frac{1-e_{\rm in}}{10^{-3}} \right).
\end{align}
\end{subequations}
The examples depicted in Figures \ref{fig:tilt_a5_I5_Om0_Lam7e3} through \ref{fig:tilt_a5_I85_Om90_Lam3e4} illustrate the qualitative features of the orbital evolution and the reliability of our estimated circularization time-scale. Note that we have included leading-order general-relativistic corrections to the secular equations as described later in this section.

For all of our examples, we adopt a truncated power-law model for the disc's surface density profile:
\begin{equation}
    \Sigma_{\rm d} = \Sigma_{1} \left( \frac{r}{r_{1}} \right)^{-\gamma}, \ \ r_{1} \leq r \leq r_{2},
\end{equation}
where $\Sigma_{1}$ is the disc's density at $r = r_{1}$. We choose $\gamma = 2$ in imitation of a viscous, gaseous accretion disc fed by the sublimation of solid grains at $r = r_{1}$ \citep*{MRB2012}. The sublimation radius $r_{\rm sub} \sim r_{1}$ is related to both the properties of the WD and the composition of the grains \citep{Rafikov2011b,Rafikov2011a}:
\begin{subequations}
\begin{align}
    r_{\rm sub} &= \frac{R_{*}}{2} \left( \frac{T_{*}}{T_{\rm sub}} \right)^{2}, \\
    &= 0.50 \RSol \left( \frac{R_{*}}{10^{-2} \RSol} \right) \left( \frac{T_{*}}{10^{4} \, {\rm K}} \right)^{2} \left( \frac{T_{\rm sub}}{10^{3} \, {\rm K}} \right)^{-2},
\end{align}
\end{subequations}
where $T_{\rm sub}$ is the sublimation temperature of grains ($\sim 1500 \, {\rm K}$ for silicates). Observations of real debris discs show that $r_{2} \sim 10 r_{1}$ in most cases \citep*{JFZ2009}. In our examples, we fix the properties of the central WD to be $M_{*} = 0.6 \MSol$, $R_{*} = 1.4 \RE$, $T_{*} = 10^{4} \, {\rm K}$. For the disc's solid component, we let $T_{\rm sub} = 1500 \, {\rm K}$, so that $r_{1} = 0.28 \RSol$. We also fix the following initial conditions of the orbit: $a_{\rm in} = 5.0 \AU$ and $r_{\rm p,in} = \RSol$ ($P_{\rm in} = 14.4 \yr$, $1 - e_{\rm in} = 9.3 \times 10^{-4}$). We vary the initial values of $I$ and $\varpi$, which specify the initial orientation of the orbit; and $\Lambda$, which fixes the ratio $\rho s / C \Sigma_{\rm d}$ at $r = r_{0}$. We halt the orbital evolution when the inclination becomes less than or equal to the disc's aspect ratio ($h/r = 10^{-3}$) because the impulse approximation is invalid for $I \lesssim h/r$.

In Figs.~\ref{fig:tilt_a5_I5_Om0_Lam7e3} through \ref{fig:tilt_a5_I85_Om90_Lam3e4}, we display the evolution of the orbital elements $a$, $e$, $I$ (as $\sin I$), and $\varpi$, as well as the characteristic orbital distances $a (1+e)$, $a (1-e)$, and $a (1-e^{2})$. By inspecting the evolution of $a (1-e^{2})$ (which is proportional to the square of the orbital angular momentum), we see that there are two distinct phases of the evolution: During the first phase, the orbit evolves essentially at constant angular momentum; neither $e$, $I$, nor $a (1-e^{2})$ changes by a discernible amount during this phase. This reflects the factor $\sim (1-e^{2})$ that appears in the expression for energy dissipation (equation \ref{eq:drag_dE/E}) but not in that for the torque (equations \ref{eq:drag_dl/l} and \ref{eq:drag_dI}). During the second phase, when $(1-e^{2})$ is of order unity, the torque exerted on the orbit by the disc becomes substantial. The quantity $a (1-e^{2})$ shrinks monotonically throughout this phase; meanwhile, both $e$ and $I$ decrease precipitously from their initial values. If we denote by $t_{I}$ the time elapsed between $t = 0$ and the instant that $I = h/r$ (so that the integration is halted artificially), then the duration of the first phase is between $80$ and $90$ per cent of $t_{I}$ in all cases we examined.

By comparing each of Figs.~\ref{fig:tilt_a5_I5_Om0_Lam7e3} through \ref{fig:tilt_a5_I85_Om90_Lam3e4}, we see that changing the initial orientation of the orbit ($I$ or $\varpi$) changes both the time-scale of evolution and the ultimate orbit of the body. For instance, by comparing Figs. \ref{fig:tilt_a5_I45_Om0_Lam7e3} and \ref{fig:tilt_a5_I45_Om45_Lam1e4} or Figs. \ref{fig:tilt_a5_I85_Om0_Lam7e3} and \ref{fig:tilt_a5_I85_Om90_Lam3e4}, we see that changing $\varpi$ can change the true circularization time-scale by factors of a few. We also see that the estimated circularization time can be either less than or greater than the true time-scale. The initial value of $I$ has a more pronounced effect on the orbital evolution, but on the whole equation (\ref{eq:tdrag}) provides a robust estimate for the migration time-scale.

A notable trend in these examples is that the radius of a circularized orbit varies systematically with the initial inclination of the orbit but not the initial apsidal angle. For low-to-moderate initial inclinations (Figs. \ref{fig:tilt_a5_I5_Om0_Lam7e3}, \ref{fig:tilt_a5_I45_Om0_Lam7e3}, and \ref{fig:tilt_a5_I45_Om45_Lam1e4}), the final orbital radius tends to be somewhat larger than the initial pericentre distance. For high initial inclinations (Figs. \ref{fig:tilt_a5_I85_Om0_Lam7e3} and \ref{fig:tilt_a5_I85_Om90_Lam3e4}), the final orbital distance can be barely half the initial pericentre distance. This reflects a significant loss of orbital angular momentum to the disc over the course of migration. It also presents the possibility of delivering planetesimals well inside of the Roche limit (as apparently is the case with J1228), even if their initial pericentre distances were outside it.

Given the body's close proximity to the WD at pericentre, one must ask whether general-relativistic corrections to the equations of motion are important. The leading-order post-Newtonian effect is prograde apsidal precession of the orbit at a rate
\begin{align} \label{eq:omega_GR}
    \dot{\varpi}_{\rm GR} &= \frac{3 G M_{*}}{c^{2} a (1-e^{2})} n, \nonumber \\
    &\simeq \frac{2 \pi}{3 \times 10^{5} \yr} \left( \frac{M_{*}}{\MSol} \right) \left( \frac{r_{\rm p}}{\RSol} \right)^{-1} \left( \frac{P}{1 \yr} \right)^{-1},
\end{align}
where in the second expression we have taken the limit $e \to 1$. Due to the definition of $\varpi$ in terms of the relative orientation between the disc and the orbit (equation \ref{eq:coordinate_vectors}), the GR contribution to $(\dif\varpi / \dif t)$ is negative:
\begin{equation}
    \frac{\dif \varpi}{\dif t} = \frac{\delta \varpi_{+} + \delta \varpi_{-}}{P} - \dot{\varpi}_{\rm GR}    
\end{equation}
Equation (\ref{eq:omega_GR}) suggests that, for an initial orbit with $P \gtrsim 1 \yr$, the precession period is comparable to the disc life-time (see Section \ref{s:drag:constraints}); this would mean that GR does not qualitatively affect the early evolution of the orbit. After the orbit has shrunk somewhat, however, the precession period becomes shorter than the migration time-scale. Thus, we have included GR effects in our examples for accurate computation of the orbital evolution. We have found that, indeed, the contribution to precession due to GR tends to overwhelm that due to drag forces in determining the overall precession rate.

\subsection{Constraints on Disc-Driven Migration} \label{s:drag:constraints}

If the candidate planetesimals around WD1145 and J1228 achieved their current orbits via inclined disc-driven migration, what constraints do they imply on the properties of the disc--planetesimal system? Clearly, we require the migration time-scale to be less than the disc life-time. Observations place the disc life-time between a few $10^{4}$ and a few $10^{6} \yr$ on average \citep{Girven+2012}, in keeping with theoretical estimates for various accretion mechanisms \citep{Rafikov2011b,Rafikov2011a,MRB2012}. We therefore adopt a fiducial disc life-time of $t_{\rm disc} = 4 \times 10^{5} \yr$ as the appropriate upper limit on the migration time-scale $t_{\rm drag}$ in this scenario. Thus, we require the dimensionless migration parameter $\Lambda$ (the number of orbits over which migration occurs; see equation \ref{eq:Lambda}) to satisfy
\begin{equation}
    \Lambda \lesssim \frac{t_{\rm disc}}{P_{\rm in}}
\end{equation}
in order for migration to occur. Unfortunately, referring to equations (\ref{eq:Lambda}), we see that the properties of the migrating planetesimal ($\rho s$) and the accretion disc ($\Sigma_{0}$) are degenerate with the initial orbital parameters ($P_{\rm in}$ and $e_{\rm in}$). For the remainder of this discussion, we will assume an initial orbital period of $10 \yr$ and an initial eccentricity of $0.999$.

In Section \ref{s:tides:constraints}, we took the (main) object in orbit around WD1145 to have $\rho s \approx 9 \times 10^{7} \, {\rm g \, cm^{-2}}$. By taking $\Lambda \lesssim 4 \times 10^{4}$ and $1 - e_{\rm in} = 10^{-3}$, we find that disc migration for this object requires a typical surface density $\Sigma \gtrsim 2.5 \, {\rm g \, cm^{-2}}$. The same reasoning applied to the J1228 system yields
\begin{equation*}
    \Sigma \gtrsim 0.2 \, {\rm g \, cm^{-2}} \left( \frac{\rho s}{8 \times 10^{6} \, {\rm g \, cm^{-2}}} \right),
\end{equation*}
where the quoted value of $\rho s$ corresponds to an iron body $10 \, {\rm km}$ in size. These results are only as certain as existing constraints on the size and mass of the orbiting objects in these systems. Nonetheless, they illustrate the minimal amount of circumstellar material required in order for disc migration to be viable. We discuss the implications of this constraint in Section \ref{s:discuss:disc}.

\section{Discussion} \label{s:discuss}

\subsection{Viscosity from Tidal Migration} \label{s:discuss:visco}

In Section \ref{s:tides:constraints}, we obtained constraints on the internal viscosity of the planetesimals in orbit around WD~1145+017 and SDSS~J1228+1040 under the assumption that their present orbits are the result of tidal evolution with Maxwell viscoelastic dissipation. Although these constraints are imprecise and are somewhat degenerate with the objects' sizes and bulk densities, they are interesting from a geophysical perspective.

\begin{figure}
    \centering
    \includegraphics[width=\columnwidth]{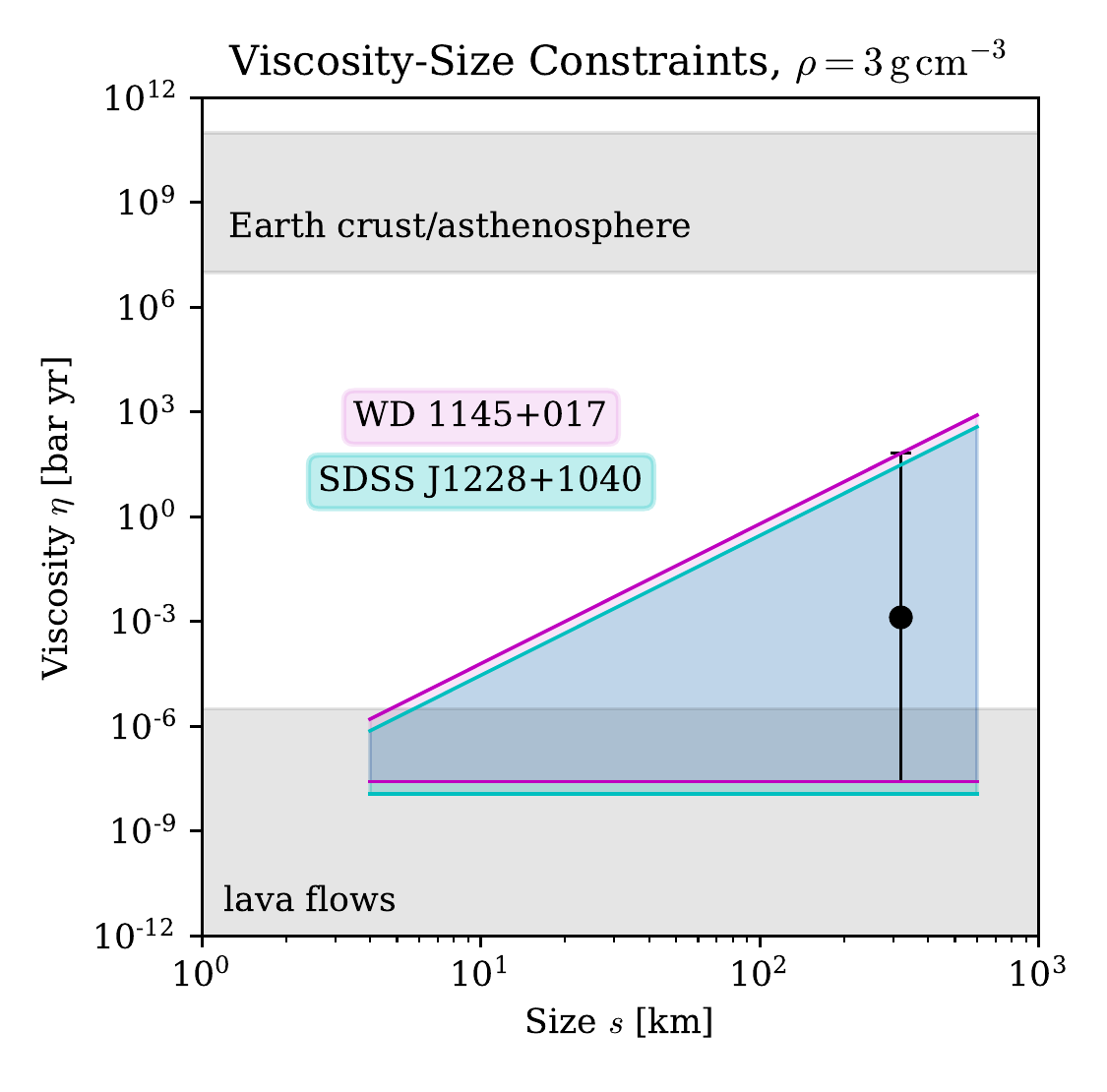}
    \caption{Maxwell viscosities inferred for WD candidate planetesimals as functions of their assumed physical sizes. The magenta and cyan shaded regions refer to the objects in orbit around WD1145 and J1228, respectively. The black point with error bars shows the result for WD1145 using the mass calculated by \citet{Rappaport+2016}. The grey shaded regions show the range of viscosity values for Earth's crust and asthenosphere (as a proxy for solid rock) and lava flows. We have fixed the bulk densities at $\rho = 3 \, {\rm g \, cm^{-2}}$ because this quantity varies by a factor of a few at most while the size can vary over several orders of magnitude.}
    \label{fig:etaphys_constraints}
\end{figure}

In Figure \ref{fig:etaphys_constraints}, we display the range of viscosity values consistent with tidal migration as a function of the assumed size of the object (setting $\rho = 3 \, {\rm g \, cm^{-3}}$ for both). We also show the typical range of viscosity values that have been measured or inferred for Earth's crust and upper mantle \citep{MF2004} and for terrestrial lava flows of various compositions at typical temperatures of $1300$--$1400 \, {\rm K}$ \citep[as compiled by][]{CPH2019}. We see that the planetesimals around WD1145 and J1228 have bulk viscosity consistent with being molten in whole or in part. On the other hand, both are inconsistent with having the same bulk viscosity as Earth's crust by orders of magnitude.

A molten interior would be consistent with the hypothesis of tidal migration, due to the large amount of orbital energy that would have to be dissipated, provided that the time-scale of tidal heating be shorter than the body's internal cooling time-scale. In a small, rocky body, the dominant cooling mechanism is thermal conduction. Using a thermal diffusivity similar to terrestrial rock ($\kappa \sim 10^{-2} \, {\rm cm^{2} \, s^{-1}}$; \citealt{TS2002}, \citealt{Hartlieb+2016}), the cooling time of a planetesimal is
\begin{equation} \label{eq:tcool}
    t_{\rm cool} \sim \frac{s^{2}}{\kappa} \approx 3 \times 10^{4} \yr \left( \frac{s}{1 \, {\rm km}} \right)^{2}.
\end{equation}

To estimate the rate of tidal heating, we use equation (\ref{eq:tides_Z1_def}) with $|Z_{1}| \sim 1$. For rock with a nominal melting point temperature $T_{\rm m} \sim 10^{3} \, {\rm K}$ and specific heat capacity $c_{P} \sim 10^{7} \, {\rm erg \, g^{-1} \, K^{-1}}$ \citep{TS2002,Hartlieb+2016}, the body is heated to the melting point on a time-scale of 
\begin{subequations} \label{eq:theat}
\begin{align}
    t_{\rm heat} &\equiv \frac{c_{P} T_{\rm m} \rho s^{3}}{|\dot{E}|} \sim \frac{c_{P} T_{\rm m} \rho r_{\rm p}^{6}}{s^{2}} \left( \frac{a^{3}}{G^{3} M_{*}^{5}} \right)^{1/2}, \\
    &\approx 7 \times 10^{4} \yr \left( \frac{c_{P} T_{\rm m}}{10^{10} \, {\rm erg \, g^{-1}}} \right) \left( \frac{r_{\rm p}}{\RSol} \right)^{6} \left( \frac{s}{1 \, {\rm km}} \right)^{-2} \nonumber \\
    & \hspace{1cm} \times \left( \frac{\rho}{1 \, {\rm g \, cm^{-3}}} \right) \left( \frac{M_{*}}{\MSol} \right)^{-5/2} \left( \frac{a}{1 \AU} \right)^{3/2}.
\end{align}
\end{subequations}
Similarly, the heat required to melt the body is $H_{\rm f} \rho s^{3}$, where $H_{\rm f} \approx 10^{10} \, {\rm erg \, g^{-1}}$ is the latent heat of fusion \citep{TS2002}. Thus, the time-scale of internal melting is
\begin{equation} \label{eq:tmelt}
    t_{\rm melt} \equiv \frac{H_{\rm f} \rho s^{3}}{|\dot{E}|} = t_{\rm heat} \left( \frac{H_{\rm f}}{c_{P} T_{\rm m}} \right).
\end{equation}
Since $H_{\rm f} \sim c_{P} T_{\rm m}$ for common minerals, heating and melting occur on roughly the same time-scale.

Comparing equations (\ref{eq:tcool}) and (\ref{eq:theat}), we see that the time-scale of heating and melting is shorter than the cooling time for bodies larger than a few km. This is the case for the planetesimals around WD1145 and J1228, and thus a molten interior is indeed plausible.

\subsection{Disc Masses from Disc Migration} \label{s:discuss:disc}

Most accretion discs around polluted WDs are composed primarily of dust rather than gas \citep{Manser+2020}. Such discs are detected mainly by their thermal emission at infrared wavelengths \citep[e.g.,][]{ZB1987,Farihi2016}; simple models of this emission can be constructed in order to estimate their radial extent and optically thin dust mass \citep[e.g.,][]{JFZB2007,JFZ2007,JFZ2009}. However, the total dust mass cannot be calculated from infrared emission alone, since the disc's infrared continuum emission may be due to an opaque component. Sometimes, it is possible to estimate the total mass of accreted material in WD's convective zone \citep{Jura2006,Zuckerman+2007}, which provides an indirect estimate of the disc's mass budget \citep{JFZ2009}.

The disc migration mechanism provides an independent estimate of the disc's total mass because the time-scale of a planetesimal's orbital evolution is determined mainly by the column-density ratio $\rho s / \Sigma_{\rm d}$. According to the constraints of Section \ref{s:drag:constraints}, and adopting a fiducial outer disc radius of $2 \RSol$, we estimate the characteristic disc mass around WD1145 to be at least
\begin{equation*}
    1.5 \times 10^{23} \, {\rm g} \left( \frac{\rho s}{9 \times 10^{7} \, {\rm g \, cm^{-3}}} \right)
\end{equation*}
and around J1228 to be at least
\begin{equation*}
    1.2 \times 10^{22} \, {\rm g} \left( \frac{\rho s}{8 \times 10^{6} \, {\rm g \, cm^{-2}}} \right).
\end{equation*}
These estimates are consistent with those given by the sources cited above.

\subsection{Caveats} \label{s:discuss:caveats}

\subsubsection{Viscoelastic Rheology} \label{s:discuss:caveats:rheology}

The validity of the constraints we have derived on the viscosity of small bodies (Section \ref{s:tides:constraints}) during tidal migration extends only as far as the validity of the Maxwell rheology in describing their response to tidal forcing. The Maxwell model is the simplest viscoelastic theory, and thus it fails to capture certain aspects of material physics. The sensitivity of tidal evolution to one's choice of rheological model has been discussed elsewhere \citep[e.g.,][]{ME2013,RH2018}. The complex Love number $\tilde{k}_{2}(\omega)$ can be calculated for other, more sophisticated rheologies \citep{RH2018}; it may be possible to extract constraints on bulk properties within these alternative models by arguments analogous to those we have used.

\subsubsection{Accretion Disc Properties} \label{s:discuss:caveats:disc}

Our calculations in Section \ref{s:drag} assume a power-law density profile for the disc \citep[see][]{MRB2012}. This is physically reasonable if the gaseous and dusty disc components coincide, as for SDSS~J1228+1040 and others \citep{Brinkworth+2009,Melis+2010}. For WD~1145+017, however, the circumstellar gas is concentrated near the sublimation radius \citep[e.g.,][]{Cauley+2018,Xu+2019a,FDX2020}, while the dust mostly coincides with transiting debris \citep[e.g.,][]{Vanderburg+2015,Xu+2019a}. Moreover, polluted WDs with detectable circumstellar gas appear to be very much the minority: a recent analysis of a joint {\it Gaia}/SDSS WD catalogue \citep{GentileFusillo+2019} found that less than $10$ per cent of dusty debris systems have gaseous components \citep{Manser+2020}. Thus, it is of interest to examine the sensitivity of the disc migration scenario to the disc's mass distribution.

We consider two simple alternatives to the power-law disc model: an uniform (or top-hat) disc with $\Sigma_{\rm d} = {\rm constant}$ and a Gaussian disc with
\begin{equation}
    \Sigma_{\rm d} \propto \exp\left[ - \frac{(r - r_{\rm ctr})^{2}}{(\Delta r)^{2}} \right].
\end{equation}
We model the latter after \citet{MRB2012}, selecting $r_{\rm ctr} = 5 r_{\rm sub}$ and $\Delta r = 0.5 r_{\rm sub}$. As before, we truncate these disc models at $r < r_{\rm sub}$ and $r > 10 r_{\rm sub}$. For the Gaussian profile, these nominal dimensions matter very little because the bulk of the disc's mass is concentrated within a few $\Delta r$ of $r_{\rm ctr}$.

We have computed the evolution of the body's orbit for the same initial conditions as shown in Figs. \ref{fig:tilt_a5_I5_Om0_Lam7e3} through \ref{fig:tilt_a5_I85_Om90_Lam3e4} using these alternative disc models. Our results for the uniform mass distribution are qualitatively and quantitatively similar to those for the power-law distribution: For sufficiently small values of $\Lambda$, the orbit can circularize and align with the accretion disc within the latter's fiducial life-time. We note in particular that the product $\Lambda P_{\rm in}$ remains a reasonable estimate of the migration time-scale; if anything, it is slightly more of an overestimate.

We obtain somewhat different results for the disc with a Gaussian mass distribution. For some orbital configurations, the orbit evolves in much the same way as for the power-law and uniform disc models. For others, migration begins normally but slows to an effective standstill while the eccentricity and inclination remain large. For still others, migration does not occur at all within the nominal disc life-time. This is mainly because a Gaussian disc can be much narrower in effect than the radial extent $r_{1} \leq r \leq r_{2}$ of the mass distribution would na\"{i}vely suggest. For $\Delta r \ll (r_{2}-r_{1})$, a disc presents a much smaller `target' for the orbit to intersect, and thus the space of orbital configurations for which disc migration is efficient is much smaller. These differences disappear when $\Delta r$ is increased by a factor of a few. Thus, we conclude that the success or failure of disc migration is sensitive mainly to the width of the disc and its total mass (which together determine the typical surface density). We reiterate that many dusty discs around polluted WDs are observed to be fairly broad \citep{JFZ2009}, which is favourable for disc migration (but see \citealt{Li+2017} for an exception).

\subsubsection{Sublimation and Ablation} \label{s:discuss:caveats:loss}

Many polluted WDs have effective temperatures in excess of $10\,000 \, {\rm K}$, with WD1145 and J1228 among them. Objects orbiting in close proximity to the WD are strongly irradiated by UV light and thus may lose mass by sublimation. This effect provides an additional constraint on the origins and properties on close-orbiting planetesimals because their hypothetical high-eccentricity migration must occur before the objects vaporize completely.

We estimate the mass-loss rate of a small body under the assumption that all of the WD's radiation that is incident on the body's geometric cross-section during a short time interval is converted to the heat required to sublime a thin layer at the surface. The resulting mass-loss rate is
\begin{equation} \label{eq:Mdot_sublim}
    \dot{M} \sim \frac{\varsigma T_{*}^{4} R_{*}^{2}}{H_{\rm s}} \frac{s^{2}}{r^{2}},
\end{equation}
where $\varsigma$ is the Stefan--Boltzmann constant, $T_{*}$ and $R_{*}$ are the WD's effective temperature and radius, $r$ is the radial distance from the WD's center, and $H_{\rm s}$ is the body's latent heat of sublimation. For a body on a highly eccentric orbit, mass loss by sublimation takes place mostly near pericentre and thus the mass lost over a single orbit is $\sim \dot{M}_{\rm p} t_{\rm p}$, where $t_{\rm p} = 1 / \Omega_{\rm p}$ is the duration of pericentre passage and $\dot{M}_{\rm p}$ is the value of equation (\ref{eq:Mdot_sublim}) at $r = r_{\rm p}$. The mass-loss rate averaged over one orbit is then
\begin{equation}
    \langle \dot{M} \rangle \sim \frac{\dot{M}_{\rm p} t_{\rm p}}{P}.
\end{equation}
Thus, the sublimation time-scale of a body with initial mass $\rho s^{3}$ is
\begin{subequations}
\begin{align}
    t_{\rm sub} &\equiv \frac{\rho s^{3}}{\langle \dot{M} \rangle} \sim \frac{\rho s H_{\rm s}}{\varsigma T_{*}^{4}} \left( \frac{r_{\rm p}}{R_{*}} \right)^{2} \left( 1-e \right)^{-3/2}, \\
    &\sim 10^{4} \yr \left( \frac{\rho}{1 \, {\rm g \, cm^{-3}}} \right) \left( \frac{s}{1 \, {\rm km}} \right) \left( \frac{H_{\rm s}}{10^{10} \, {\rm erg \, g^{-1}}} \right) \nonumber \\ &\times \left( \frac{r_{\rm p}}{\RSol} \right)^{2} \left( \frac{1-e}{10^{-3}} \right)^{-3/2} \left( \frac{T_{*}}{10^{4} \, {\rm K}} \right)^{-4} \left( \frac{R_{*}}{10^{-2} \RSol} \right)^{-2}.
\end{align}
\end{subequations}

When sublimation is taken into account, successful high-eccentricity migration requires $t_{\rm circ} < t_{\rm sub}$ for the migration mechanism in question. For tidal migration, this can present a more stringent constraint on a planetesimal's properties than the condition $t_{\rm circ} < t_{\rm WD}$. For instance, the life-time of the object in orbit around WD1145 (assuming $\rho = 3 \, {\rm g \, cm^{-3}}$ and $s = 320 \, {\rm km}$) would be between $1$ and $10 \Myr$ for $H_{\rm s}$ between $10^{10}$ and $10^{11} \, {\rm erg \, g^{-1}}$, significantly shorter than the WD cooling age of $\sim 200 \Myr$. Under this alternative condition, we calculate a more stringent constraint of $Z_{1} \gtrsim 8.4 \times 10^{-4}$.

On the other hand, the sublimation time-scale for rocky bodies at least a few km in size is at least as long as the fiducial life-time of accretion discs around polluted WDs, a few $10^{5} \yr$. Moreover, the life-time of such an object can be extended by the disc's shielding effect if the object's orbit becomes embedded within it. These considerations suggest that our earlier constraints on disc migration are mostly insensitive to this additional effect.

Planetesimals undergoing inclined disc migration might also lose significant mass by ablation, owing to the supersonic relative velocity between the body and the disc. We estimate the ablation time-scale following \citet{Jura2008}. During each pericenter passage, we suppose that ablation removes a layer from the planetsimal's surface of thickness
\begin{equation}
    \delta s \sim \frac{\Sigma_{\rm d}}{\rho} \frac{y}{\sin I},
\end{equation}
where $y \approx 0.1$ is the sputtering yield. The ablation time-scale is then
\begin{equation}
    t_{\rm abl} \equiv \left( \frac{s}{\delta s} \right) P \sim \frac{\rho s}{\Sigma_{\rm d}} \frac{\sin I}{y} P.
\end{equation}
Comparing this expression to the characteristic circularization time under drag forces (equation \ref{eq:tdrag}), we have
\begin{equation}
    \frac{t_{\rm abl}}{t_{\rm drag}} \sim \frac{\sin I}{y (1 - e)}.
\end{equation}
Since we take the initial orbit to be highly eccentric and moderately-to-highly inclined, we see that $t_{\rm abl} \gg t_{\rm drag}$. This indicates that ablation is not important in the early stages of orbital evolution. As the orbit circularizes and becomes aligned with the disc over time, the ablation time-scale may become comparable to the drag time-scale.

On the whole, it is plausible that the objects orbiting WD1145 and J1228 have lost mass in the course of their migration by either sublimation or ablation. If so, a consistent theory of their origin must take these effects into account \citep*[see also][]{VEG2015}.

\section{Summary} \label{s:summary}

We have studied the high-eccentricity migration of the candidate planetesimal companions to WD~1145+017 and SDSS~J1228+1040 from an initial orbit at several au to their present locations near or within the tidal disruption radius. We have shown that either internal tidal dissipation or drag forces from an accretion disc could be responsible for the circularization on an appropriate time-scale. We have presented general analytical expressions for the migration rates that can be easily rescaled to various situations and parameters. Our principal conclusions regarding each mechanism are as follows:
\begin{itemize}
    \item If tidal friction is responsible for the migration of the candidate planetesimals, then it is possible to constrain both their tensile strength and their internal viscosity. The requisite tensile strength to avoid disruption can be slightly high compared to those of silicate meteorites. Under a Maxwell rheology, the range of viscosity that allows tidal migration within the WD cooling age is inconsistent with the viscosity of Earth's crust or mantle, but includes a wide range of measured values for molten rock. We speculate that the objects may have been partly or totally molten due to internal heating during their migration.
    \item If drag forces from an accretion disc around the WD is the main migration mechanism, then it is possible constrain the column density of the disc relative to the density and size of the migrating body. Accordingly, we have obtained the characteristic total disc mass required for complete migration within a disc life-time of several $10^{5} \yr$. The results for WD1145 and J1228 are consistent with previous estimates of the total mass of metals accreted by other polluted WDs.
\end{itemize}

Is either scenario favoured over the other in explaining the origin of the candidate planetesimals? Tidal migration seems the most natural explanation by default, given that debris from tidal disruption of asteroids seems to be ubiquitous around polluted WDs -- clearly, it is possible to deliver planetesimals to pericentre distances near or inside the Roche radius. In that case, the most likely reason that the observed objects were not fully disrupted is that they are monoliths, rather than rubble piles, and therefore have a moderate amount of internal strength.

While the disc migration scenario is somewhat non-standard, it is physically well motivated in that it relies on little more than the presence of a sufficiently massive accretion disc when a `fresh' planetesimal approaches the tidal radius for the first time. Because massive discs are relatively rare and last for finite time, the limiting factor is the rate at which small bodies are excited onto highly eccentric orbits, which depends on the presence and orbital configuration of outer planets or binary stellar companions.

\subsubsection*{Data availability}

The data underlying this article will be shared on reasonable request to the corresponding author.

\section*{Acknowledgements}

We thank Yubo Su, Michelle Vick, Dimitri Veras, and Brian Metzger for helpful discussions. We also thank the referee, Alexander Mustill, for insightful comments that improved the manuscript. DL thanks the Department of Astronomy and the Miller Institute for Basic Science at the University of California, Berkeley, for hospitality while part of this work 
was carried out. This work has been supported in part by NSF grant AST-1715246 and NASA grant 80NSSC19K0444.

{\it Software:} matplotlib \citep{Hunter2007}, SciPy \citep{Virtanen+2020}

\bibliography{asteroid}
\bibliographystyle{mnras}

\appendix

\section{Jagged and Smooth Regimes of Viscoelastic Tidal Dissipation} \label{app:viscotides}

In Section \ref{s:tides:Zstuff}, we identified two qualitative behaviours of the functions $Z_{1}$ and $Z_{2}$ for viscoelastic tidal dissipation, which we dubbed ``jagged'' and ``smooth.'' In this Appendix, we elaborate on the mathematical origin of these r\'{e}gimes.

\begin{figure}
    \centering
    \includegraphics[width=\columnwidth]{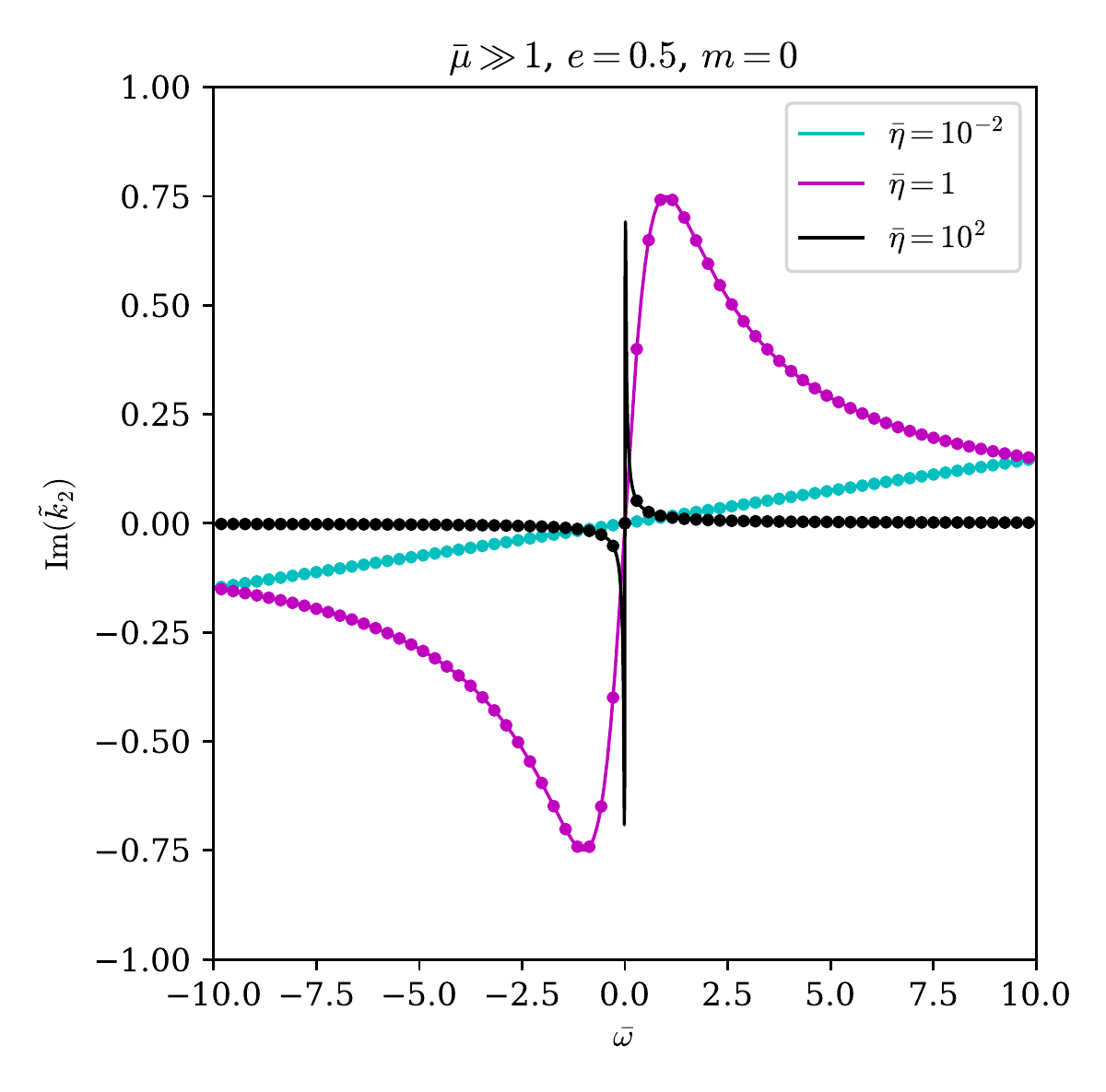}
    \caption{The solid curves illustrate $\Im(\tilde{k}_{2})$ as a function of the dimensionless tidal forcing frequency $\bar{\omega}$ for three dimensionless viscosities, negligible self-gravity, and fixed $e = 0.5$. The points marked along each curve indicate the values of $\Im(\tilde{k}_{2})$ at each forcing frequency for modes with $m = 0$.}
    \label{fig:k2samples_e50}
\end{figure}

Recall that the imaginary part of the Love number for a mode with (dimensionless) forcing frequency $\bar{\omega}$ is
\begin{equation}
    \Im(\tilde{k}_{2}) \simeq \frac{3}{2} \bar{\omega} \bar{\eta} \left[ 1 + \left( \frac{\bar{\omega} \bar{\eta}}{\bar{\mu}} \right)^{2} (1 + \bar{\mu})^{2} \right]^{-1}.
\end{equation}
In Figure \ref{fig:k2samples_e50}, we illustrate this function for several combinations of $\bar{\mu}$ and $\bar{\eta}$. Its profile has odd symmetry, with two `humps' peaked at
\begin{equation}
    \bar{\omega}_{\rm pk} = \pm \frac{(\bar{\mu}/\bar{\eta})}{1 + \bar{\mu}}.
\end{equation}
This frequency is determined by competition between a body's elastic ($\bar{\mu}$) and viscous ($\bar{\eta}$) responses to tidal forcing and is analogous to the resonant frequency of a damped, driven harmonic oscillator. The width $\Delta$ of the resonant `humps' is of the same order of magnitude as $\bar{\omega}_{\rm pk}$; in the case $\bar{\mu} \gg 1$, which is most relevant for our purposes in the main text, we have simply $\Delta \sim 1/\bar{\eta}$.

The functions $Z_{1}$ and $Z_{2}$ are evaluated by summing the value of $\Im(\tilde{k}_{2})$ (weighted by the square of the Hansen coefficient) over the set of dimensionless forcing frequencies
\begin{equation}
    \bar{\omega}_{mN} = \frac{N n - m \Omega_{\rm s}}{\Omega_{\rm p}}.
\end{equation}
We observe that, for a fixed value of $m$, the frequencies of modes $N$ and $(N+1)$ are spaced by $n/\Omega_{\rm p} \sim (1-e)^{-3/2}$. The transition between the smooth and jagged behaviours of $Z_{1}$ and $Z_{2}$ is determined by the relative values of the viscoelastic resonance width $\Delta$ and the mode spacing $n/\Omega_{\rm p}$. For $\Delta \gg n/\Omega_{\rm p}$, the set of forcing frequencies samples the resonant hump very finely, so that the series expressions for $Z_{1}$ and $Z_{2}$ approximate definite integrals of $\Im(\tilde{k}_{2})$ with respect to $\bar{\omega}$. This corresponds to the smooth r\'{e}gime. On the other hand, when $\Delta \lesssim n/\Omega_{\rm p}$, only a few modes (if any) coincide with the `humps' of $\Im(\tilde{k}_{2})$. Thus, the values of $Z_{1}$ and $Z_{2}$ are sensitive to small changes in the values of the forcing frequencies, giving rise to the observed jagged behaviour.

This behaviour can be understood visually by marking the points sampled by $\bar{\omega}_{mN}$ along the profile of $\Im(\tilde{k}_{2})$, as we have done in Figure \ref{fig:k2samples_e50}. For the cases with $\bar{\eta} = 1$ and $\bar{\eta} = 10^{-2}$, the resonant `humps' are sufficiently broad to encompass multiple forcing frequencies; as a result, their forcing functions are smooth with respect to $\Omega_{\rm s}/\Omega_{\rm p}$. On the other hand, in the case $\bar{\eta} = 10^{2}$ the profile of $\Im(\tilde{k}_{2})$ is so narrow that the forcing frequencies ``ignore'' or ``pass by'' the resonant region. A nonzero rotation rate would shift the set of forcing frequencies (for $m = \pm 2$ modes), so that overlap with the resonant region becomes possible.

\section{Coplanar Disc Migration} \label{app:coplanar}

In Section \ref{s:drag:impulse}, we calculated the orbital evolution of a planetesimal intersecting a circumstellar disc at a relatively high inclination ($I \gtrsim h/r$), in which case the impulse approximation is valid. In this Appendix, we discuss the case $I \lesssim h/r$, where the planetesimal's orbit and the disc are nearly coplanar.

As before, we consider the planetesimal's orbit to be Keplerian and we take the drag force per unit mass to be
\begin{equation}
    \mathbf{f} = - \frac{C}{4 h} \frac{\Sigma_{\rm d}}{\rho s} |\Delta \mathbf{v}| \Delta\mathbf{v}
\end{equation}
where the notations are the same as in the main text (see equation \ref{eq:fdrag_spec}). This expression is valid when the relative velocity is supersonic or has a high Reynolds number and when $s < h$.

The changes of the orbital energy and angular momentum per orbit are given by
\begin{subequations}
\label{eq:coplanar_dX}
\begin{align}
    \delta\mathcal{E} &= - \frac{C}{4 \rho s} \oint \frac{\Sigma_{\rm d}(r)}{h(r)} |\Delta\mathbf{v}| (\mathbf{v} \cdot \Delta\mathbf{v}) \, \dif t, \label{eq:coplanar_dE} \\
    \delta\ell &= - \frac{C}{4 \rho s} \oint \frac{\Sigma_{\rm d}(r)}{h(r)} |\Delta\mathbf{v}| \left[ (\mathbf{r} \times \Delta\mathbf{v}) \cdot \lhat \right] \dif t, \label{eq:coplanar_dl}
\end{align}
\end{subequations}
where $\Sigma_{\rm d}$ and $h(r)$ are the disc's density profile and vertical scale-height evaluated at the (time-dependent) position of the planetesimal.

\begin{figure}
    \centering
    \includegraphics[width=\columnwidth]{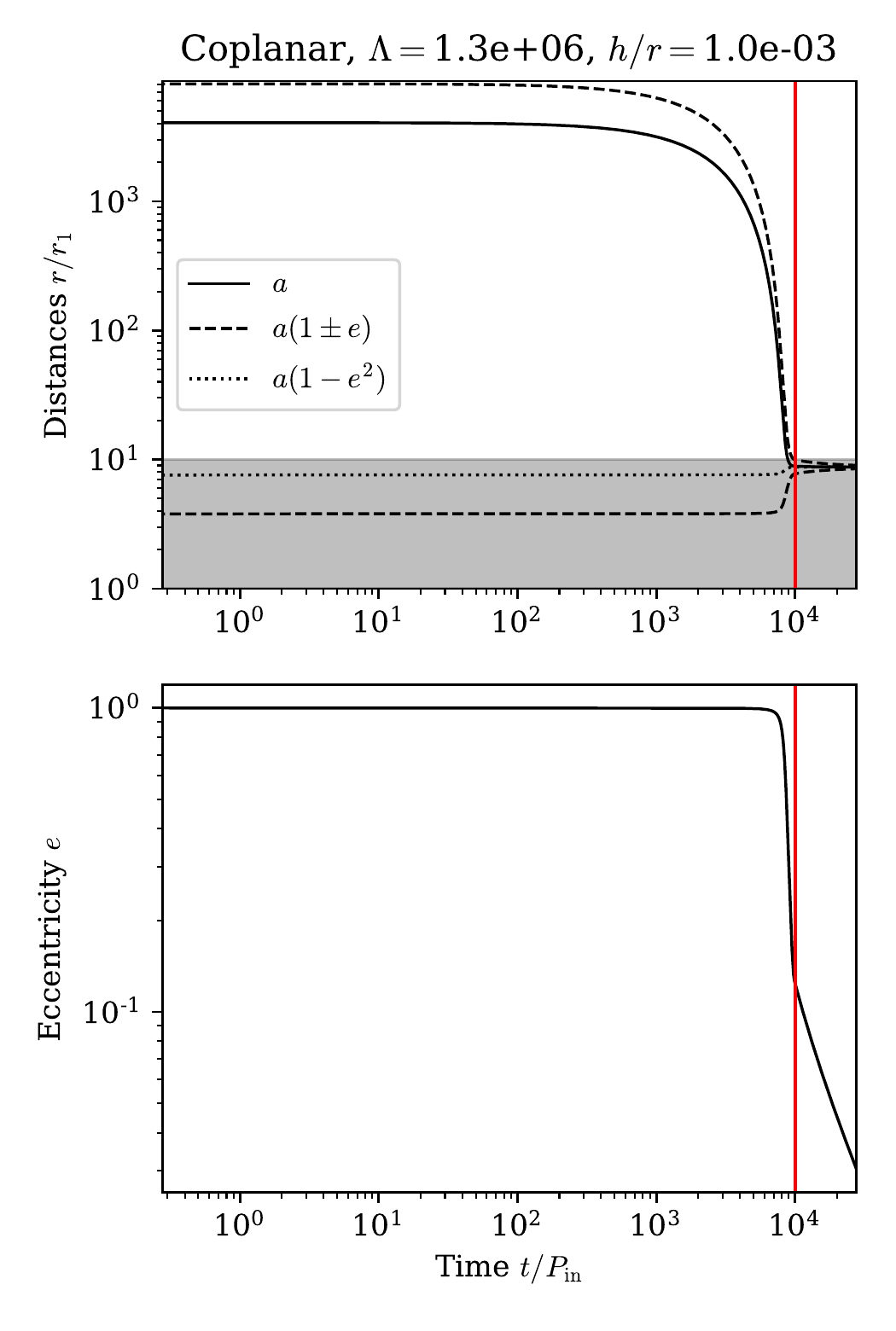}
    \caption{Orbital evolution of a planetesimal due to drag forces in a coplanar disc. The migration parameter is $\Lambda \approx 1.3 \times 10^{6}$ and the disc aspect ratio is $h/r = 10^{-3}$. The epoch $t = 7.5 \Lambda (h/r) P_{\rm in}$ is marked with a vertical red line. Upper panel: Characteristic orbital distances $a$ (solid black curve), $a (1 \pm e)$ (dashed), and $a (1-e^{2})$ (dotted) expressed in units of the disc's inner radius $r_{1}$. The shaded region indicates the radial extent of the circumstellar disc. Lower panel: Eccentricity $e$.}
    \label{fig:coplanar_drag}
\end{figure}

In Figure \ref{fig:coplanar_drag}, we show an example of coplanar orbital evolution from a similar initial condition ($a,e$) to the cases shown in Figs. \ref{fig:tilt_a5_I5_Om0_Lam7e3} through \ref{fig:tilt_a5_I85_Om90_Lam3e4} and with the power-law mass distribution from Section \ref{s:drag}. As then, we find that the circularization time-scale can be estimated simply for initial orbits with $e \to 1$:
\begin{equation} \label{eq:ta_coplanar}
    t_{\rm drag} \approx 7.5 \Lambda \left( \frac{h}{r} \right) P_{\rm in},
\end{equation}
where $\Lambda$ is the same as in equation (\ref{eq:Lambda}) except that the column density $\Sigma_{\rm d}$ is evaluated at $r = r_{\rm p}$. Comparing equations (\ref{eq:ta_coplanar}) and (\ref{eq:tdrag}), we see that coplanar disc migration can be orders of magnitude faster than inclined disc migration for a thin disc, all else being equal.

Several features of coplanar orbital evolution differ qualitatively from the inclined case. The most important of these is that, depending on the relative sizes of the disc and orbit, the torque exerted by drag can be positive. This aspect can be explained in terms of the kinematics of Keplerian orbits. The sign of the torque exerted by drag on the orbit at a distance $r$ from the star is determined by
\begin{equation}
    \mathcal{N} \propto - \left[ (\mathbf{v} - \mathbf{v}_{\rm d}) \cdot \phihat \right],
\end{equation}
where $\phihat$ is the unit vector in the azimuthal direction. This implies that the instantaneous torque is negative when the body has a greater tangential velocity than the disc (when the body feels a headwind) and positive in the opposite case (a tailwind). Since the disc and body are assumed to move (locally) on circular and eccentric Keplerian orbits, respectively, it can be shown that the sign of the instantaneous torque is
\begin{equation}
    {\rm sgn}(\mathcal{N}) =
    \left\{
    \begin{array}{ll}
        +1, & r > a (1-e^{2}); \\
        -1, & r < a (1-e^{2}); \\
        0, & r = a (1-e^{2}).
    \end{array}
    \right.
\end{equation}
The sign of the total torque is determined by the definite integral in equation (\ref{eq:coplanar_dl}).

\label{lastpage}

\end{document}